\documentclass[11pt,a4paper]{article}

\usepackage{amssymb}
\usepackage[dvips]{graphicx}
\usepackage{bm}

\begin{document}

\title{New form of the exact NSVZ $\beta$-function: the three-loop verification for terms containing Yukawa couplings}

\author{A.E.~Kazantsev, V.Yu.~Shakhmanov, K.V.~Stepanyantz\\
{\small{\em Moscow State University}}, {\small{\em  Physical
Faculty, Department  of Theoretical Physics}}\\
{\small{\em 119991, Moscow, Russia}}}

\maketitle

\begin{abstract}
We investigate a recently proposed new form of the exact NSVZ $\beta$-function, which relates the $\beta$-function to the anomalous dimensions of the quantum gauge superfield, of the Faddeev--Popov ghosts, and of the chiral matter superfields. Namely, for the general renormalizable ${\cal N}=1$ supersymmetric gauge theory, regularized by higher covariant derivatives, the sum of all three-loop contributions to the $\beta$-function containing the Yukawa couplings is compared with the corresponding two-loop contributions to the anomalous dimensions of the quantum superfields. It is demonstrated that for the considered terms both new and original forms of the NSVZ relation are valid independently of the subtraction scheme if the renormalization group functions are defined in terms of the bare couplings. This result is obtained from the equality relating the loop integrals, which, in turn, follows from the factorization of the integrals for the $\beta$-function into integrals of double total derivatives. For the renormalization group functions defined in terms of the renormalized couplings we verify that the NSVZ scheme is obtained with the higher covariant derivative regularization supplemented by the subtraction scheme in which only powers of $\ln\Lambda/\mu$ are included into the renormalization constants.
\end{abstract}

\section{Introduction}
\hspace*{\parindent}

The $\beta$-function of ${\cal N}=1$ supersymmetric gauge theories is related to the anomalous dimensions of the chiral matter superfields by the equation

\begin{equation}\label{NSVZ}
\beta(\alpha,\lambda) = -\frac{\alpha^2(3C_{2}-T(R)+C(R)_{i}{}^{j} (\gamma_\phi)_j{}^i(\alpha,\lambda)/r)}{2\pi(1-C_{2}\alpha/(2\pi))},
\end{equation}

\noindent
which is usually called the exact Novikov--Shifman--Vainshtein--Zakharov (NSVZ) $\beta$-function \cite{Novikov:1983uc,Jones:1983ip,Novikov:1985rd}. In our notation $r$ is the dimension of the gauge group,

\begin{equation}
\mbox{tr}(T^{A}T^{B}) = T(R) \delta^{AB},\qquad C(R)_i{}^j = (T^{A}T^{A})_i{}^j,\qquad C_{2}\delta^{CD}=f^{ABC}f^{ABD},
\end{equation}

\noindent
where $T^A$ are the generators of the representation in which the matter superfields lie and $f^{ABC}$ are the structure constants.

For the ${\cal N}=1$ supersymmetric Yang--Mills (SYM) theory without chiral matter Eq. (\ref{NSVZ}) gives the exact expression for the $\beta$-function in the form of a geometric series. The NSVZ equation can be used for proving the finiteness of the ${\cal N}=2$ supersymmetric gauge theories beyond the one-loop approximation \cite{Grisaru:1982zh,Howe:1983sr,Buchbinder:1997ib}, which follows from the NSVZ relation provided that the quantization procedure does not  break ${\cal N}=2$
supersymmetry \cite{Shifman:1999mv,Buchbinder:2014wra,Buchbinder:2015eva}. The finiteness of ${\cal N}=4$ SYM theory \cite{Grisaru:1982zh,Howe:1983sr,Mandelstam:1982cb,Brink:1982pd} also follows from the exact NSVZ $\beta$-function.

As a rule, various derivations of Eq. (\ref{NSVZ}) are based on some general arguments such as the structure of instanton contributions \cite{Novikov:1983uc, Novikov:1985rd,Shifman:1999mv}, anomalies \cite{Jones:1983ip,Shifman:1986zi,ArkaniHamed:1997mj}, and the non-renormalization of the topological term
\cite{Kraus:2002nu}. However, obtaining the exact NSVZ $\beta$-function directly, by using the tools of the perturbation theory, appears to be a very complicated problem. For ${\cal N}=1$ supersymmetric theories regularized by dimensional reduction \cite{Siegel:1979wq} in the modified minimal subtraction scheme \cite{Bardeen:1978yd} explicit three- and four-loop calculations have been done in Refs. \cite{Avdeev:1981ew,Jack:1996vg,Jack:1996cn,Jack:1998uj} and \cite{Harlander:2006xq,Jack:2007ni}, respectively. These calculations demonstrate that for this renormalization prescription (which is usually called ``the $\overline{\mbox{DR}}$ scheme'') the NSVZ equation (\ref{NSVZ}) is not valid starting from three loops. This occurs due to the scheme-dependence of the NSVZ relation. The general equations describing how the NSVZ relation transforms under finite renormalizations can be found in \cite{Kutasov:2004xu,Kataev:2014gxa}. From these equations one can see that Eq. (\ref{NSVZ}) is valid only in certain (NSVZ) subtraction schemes. According to \cite{Jack:1996vg,Jack:1996cn, Jack:1998uj,Harlander:2006xq}, up to the four-loop order, the NSVZ scheme is related to the $\overline{\mbox{DR}}$ scheme by a finite renormalization of the gauge coupling constant. Taking into account the presence of various group theory factors, the existence of this finite renormalization appears to be highly non-trivial. However, at present, there is no general prescription on how to obtain the NSVZ scheme in all orders if a theory is regularized by dimensional reduction.

The situation changes significantly in the case of using the Slavnov higher covariant derivative regularization. This regularization has first been introduced for non-supersymmetric gauge theories in \cite{Slavnov:1971aw,Slavnov:1972sq}. In the supersymmetric case it can be formulated in terms of ${\cal N}=1$ superfields  \cite{Krivoshchekov:1978xg, West:1985jx}, so that it allows calculating quantum corrections in a manifestly supersymmetric way. (Note that dimensional reduction is not self-consistent \cite{Siegel:1980qs} and, in principle, can break supersymmetry in higher orders \cite{Avdeev:1981vf,Avdeev:1982np,Avdeev:1982xy}.) A manifestly ${\cal N}=2$ version of the higher covariant derivative regularization is also known \cite{Buchbinder:2015eva}.

To explain why using the higher derivative regularization naturally leads to the NSVZ $\beta$-function, we first note that, according to Refs. \cite{Kataev:2013eta,Kataev:2013csa}, one should distinguish between renormalization group functions (RGFs) defined in terms of the bare couplings and RGFs defined in terms of the renormalized couplings. The former RGFs are independent of the renormalization prescription, but depend on the regularization. It is possible to demonstrate that with dimensional reduction they do not satisfy the NSVZ equation, see \cite{Aleshin:2015qqc,Aleshin:2016rrr}. However, if an ${\cal N}=1$ supersymmetric gauge theory is regularized by higher derivatives, then, at least in the Abelian case, the NSVZ relation is valid for RGFs defined in terms of the bare coupling constants independently of the subtraction scheme \cite{Stepanyantz:2011jy,Stepanyantz:2014ima}. This fact is based on the observation that the loop integrals giving the $\beta$-function defined in terms of the bare couplings turn out to be integrals of total derivatives in the momentum space \cite{Soloshenko:2003nc}. Later it was noted that they are also integrals of double total derivatives \cite{Smilga:2004zr}. Integrating the total derivative it is possible to reduce the number of loop integrals by~1. In the Abelian case this allows relating the $L$-loop contribution to the $\beta$-function to the $(L-1)$-loop contribution to the anomalous dimension of the chiral matter superfields and obtaining the NSVZ $\beta$-function in all orders \cite{Stepanyantz:2011jy,Stepanyantz:2014ima}.

It is interesting that in the Abelian case the relation between the $\beta$-function and the anomalous dimension of the matter superfields \cite{Vainshtein:1986ja,Shifman:1985fi} has a simple graphical interpretation. As a starting point, we consider an $L$-loop graph without external lines. The $L$-loop contribution to the $\beta$-function which corresponds to this graph is obtained by summing all superdiagrams obtained by attaching to it two external lines of the gauge superfield in all possible ways. This sum appears to be an integral of a double total derivative. On the other hand, the $(L-1)$-loop contribution to the anomalous dimension comes from 1PI superdiagrams obtained by cutting matter lines in the considered graph in all possible ways. Taking the integral of the total derivative we relate these two contributions. The three-loop calculation which illustrates this reasoning can be found in \cite{Kazantsev:2014yna}.

Note that a similar factorization into integrals of double total derivatives also takes place for some other theories, regularized by higher derivatives, and also leads to NSVZ-like relations. For example, it allows deriving the NSVZ-like expressions for the Adler $D$-function \cite{Adler:1974gd} in ${\cal N}=1$ SQCD \cite{Shifman:2014cya,
Shifman:2015doa} or for the anomalous dimension of the photino mass in softly broken ${\cal N}=1$ SQED \cite{Nartsev:2016nym}. (NSVZ-like relations for the renormalization of the gaugino mass in theories with softly broken supersymmetry have first been constructed in \cite{Hisano:1997ua,Jack:1997pa,Avdeev:1997vx}.) Also there are a lot of calculations (see, e.g., \cite{Pimenov:2009hv,Stepanyantz:2011zz,Stepanyantz:2011bz,Shakhmanov:2017soc}) pointing out that the factorization of loop integrals into integrals of double total derivatives is valid for the ${\cal N}=1$ supersymmetric non-Abelian gauge theories. However, in this case it is not so easy to explain graphically the origin of Eq. (\ref{NSVZ}), for example, due to the denominator depending on the gauge coupling constant. Moreover, cutting the lines of quantum superfields one obtains anomalous dimensions not only of the chiral matter superfields, but also of the quantum gauge superfield and of the Faddeev--Popov ghosts. However, both these problems are overcome by the help of the non-renormalization theorem for the triple gauge-ghost vertices derived in \cite{Stepanyantz:2016gtk}. The finiteness of the vertices with a single leg of the quantum gauge superfield and two ghost legs allows rewriting the exact NSVZ $\beta$-function (\ref{NSVZ}) in a new equivalent form

\begin{equation}\label{RewrittenNSVZ}
\frac{\beta(\alpha,\lambda)}{\alpha^2} = -\frac{1}{2\pi}\Big(3C_2 -T(R) -2C_2 \gamma_c(\alpha,\lambda) -2C_2\gamma_V(\alpha,\lambda) +C(R)_i{}^{j}(\gamma_\phi)_j{}^i(\alpha,\lambda)/r\Big).
\end{equation}

\noindent
This equation relates the $\beta$-function to the anomalous dimensions of the quantum superfields, namely, of the Faddeev--Popov ghosts, of the quantum gauge superfield, and of the chiral matter superfields. It does not contain the denominator depending on the gauge coupling constant and admits the same graphical interpretation as in the Abelian case. So, it is reasonable to assume that it is Eq.~(\ref{RewrittenNSVZ}) that appears in the perturbative calculations made with the higher derivative regularization. This conjecture has been confirmed by comparing the two-loop $\beta$-function with the one-loop anomalous dimensions in Ref. \cite{Shakhmanov:2017wji}.\footnote{To simplify the calculations, in Ref. \cite{Shakhmanov:2017wji} the theory was regularized by the BRST non-invariant version of the higher derivative method supplemented by a special subtraction scheme proposed in Ref. \cite{Slavnov:2003cx} which restores the Slavnov--Taylor identities.} However, the NSVZ equation can be non-trivially checked only by comparing the three-loop $\beta$-function with the two-loop anomalous dimensions, because only in this approximation the scheme-dependence becomes essential. Such a verification has been done in \cite{Shakhmanov:2017soc} for terms quartic in the Yukawa couplings, where it was confirmed that Eq. (\ref{RewrittenNSVZ}) is valid for RGFs defined in terms of the bare couplings and has the same graphical interpretation as in the Abelian case. However, for such terms there is no essential difference between Eqs.~(\ref{NSVZ}) and (\ref{RewrittenNSVZ}). That is why it is desirable to verify Eq. (\ref{RewrittenNSVZ}) for other terms. The complete three-loop calculation is very complicated and here we calculate only terms containing the Yukawa couplings including the ones proportional to $\alpha\lambda^2$. This allows to verify non-trivially the term in Eq. (\ref{RewrittenNSVZ}) which contains the anomalous dimension of the quantum gauge superfield.

Note that so far we have been discussing RGFs defined in terms of bare couplings which are scheme independent for a fixed regularization. However, standardly, RGFs are defined in terms of the renormalized couplings and depend on the subtraction scheme, see, e.g., \cite{Vladimirov:1975mx}. As we have already mentioned, they satisfy Eq. (\ref{NSVZ}) (or Eq. (\ref{RewrittenNSVZ})) only in the NSVZ schemes. If we know that RGFs defined in terms of the bare couplings satisfy the NSVZ equation with the higher derivative regularization, then the NSVZ scheme can be constructed in all loops by the help of a very simple prescription. In the Abelian case this prescription has been constructed in \cite{Kataev:2013eta,Kataev:2013csa}, see also Ref. \cite{Kataev:2014gxa} for a brief review. The main idea of these papers is that RGFs defined in terms of the bare coupling constant $\alpha_0$ coincide with the ones defined in terms of the renormalized coupling constant $\alpha$ after a formal replacement $\alpha_0\to \alpha$, if the renormalization constants satisfy the conditions

\begin{equation}\label{ConditionSQED}
Z(\alpha, x_{0})=1,\qquad Z_{3}(\alpha,x_{0})=1.
\end{equation}

\noindent
Here $x_0$ is a fixed value of $x = \ln(\Lambda/\mu)$, $\Lambda$ is the parameter in the higher derivative term with the dimension of mass, and $\mu$ is the renormalization point. In the case of using the higher derivative regularization RGFs defined in terms of the bare coupling constant satisfy the NSVZ relation. Therefore, it is also satisfied by RGFs defined in terms of the renormalized couplings under the conditions (\ref{ConditionSQED}). This implies that in the Abelian case the boundary conditions (\ref{ConditionSQED}) give the NSVZ scheme in all orders. Note that for $x_0 = 0$ from Eq. (\ref{ConditionSQED}) we obtain that only powers of $\ln\Lambda/\mu$ should be included into the renormalization constants. Following Ref. \cite{Shakhmanov:2017wji} (see also Ref. \cite{Stepanyantz:2017sqg} for a more detailed explanation) we will call this renormalization prescription ``$\mbox{HD}+\mbox{MSL}$'', where HD means that the theory should be regularized by higher derivatives and MSL is the abbreviation for minimal subtractions of logarithms. Note that $\mbox{HD}+\mbox{MSL}$ also gives the NSVZ-like schemes for the Adler $D$-function in ${\cal N}=1$ SQCD \cite{Kataev:2017qvk} and for the anomalous dimension of the photino mass in softly broken ${\cal N}=1$ SQED \cite{Nartsev:2016mvn}. It was suggested in Ref. \cite{Stepanyantz:2016gtk} that the NSVZ scheme for non-Abelian ${\cal N}=1$ supersymmetric gauge theories is also obtained by the $\mbox{HD}+\mbox{MSL}$ prescription. In this paper we will check this statement by explicit calculation.

The paper is organized as follows: In Sect. \ref{SectionHigherDerivatives} we describe the considered theory and its regularization by the higher covariant derivative method. A part of the three-loop contribution to the $\beta$-function containing the Yukawa couplings is calculated in Sect. \ref{SectionThreeLoopBeta}. In particular, we demonstrate that it is given by integrals of double total derivatives in the momentum space, so that one of the momentum integrals can be calculated analytically. In Sect. \ref{SectionBareNSVZ} the considered part of the $\beta$-function is compared with the corresponding (two-loop) contributions to the anomalous dimensions of the quantum gauge superfield and of the matter superfields. In particular, we demonstrate that the NSVZ equation in the form (\ref{RewrittenNSVZ}) is really valid for RGFs defined in terms of the bare couplings due to the equality relating the corresponding loop integrals. In Sect. \ref{SectionBareRGFs} we obtain explicit expressions for RGFs defined in terms of the bare couplings by calculating the corresponding loop integrals. RGFs defined in terms of the renormalized couplings are found in Sect. \ref{SectionRenormalizedRGFs}. In Sect. \ref{SectionNSVZScheme} we demonstrate that these RGFs satisfy the NSVZ relation both in the form (\ref{RewrittenNSVZ}) and in the form (\ref{NSVZ}) in the $\mbox{HD}+\mbox{MSL}$ renormalization scheme.

\section{Higher covariant derivative regularization for ${\cal N}=1$ supersymmetric theories}
\hspace*{\parindent}\label{SectionHigherDerivatives}

In this paper we investigate quantum corrections in ${\cal N}=1$ supersymmetric theories. It is convenient to describe such theories by the help of ${\cal N}=1$ superspace with the coordinates $(x^\mu,\theta)$, because in this case ${\cal N}=1$ supersymmetry is a manifest symmetry, see, e.g., \cite{Gates:1983nr,West:1990tg,Buchbinder:1998qv}. In this formalism the gauge field is included into the Hermitian gauge superfield $V$. The corresponding gauge superfield strength is described by the chiral Weyl spinor superfield

\begin{equation}
W_a = \frac{1}{8} \bar{D}^2(e^{-2V}D_a e^{2V}).
\end{equation}

\noindent
The chiral matter superfields $\phi_i$ belong to the representation $R$ of the gauge group $G$. Then in the massless limit the general renormalizable ${\cal N}=1$ gauge theory with a simple gauge group $G$ and chiral matter in the representation $R$ is described by the action

\begin{equation}\label{Action}
S=\frac{1}{2e_0^2}\mbox{tr Re}\int d^4x\,d^2\theta\, W^a W_a + \frac{1}{4}\int d^4x\, d^4\theta\, \phi^{*i} (e^{2V})_i{}^j \phi_j + \left(\frac{1}{6}\int d^4x\,d^2\theta\, \lambda_0^{ijk}\phi_i\phi_j\phi_k + \mbox{c.c.}\right).
\end{equation}

\noindent
(Note that in our notation the subscript $0$ denotes bare couplings.) In the first term the gauge superfield $V$ inside $W_a$ can be written as $V = e_0 V^A t^A$, where $t^A$ are the generators of the fundamental representation normalized by the condition $\mbox{tr}(t^A t^B) = \delta^{AB}/2$. However, in the second term (which contains the chiral superfields $\phi$) $V= e_0 V^A T^A$, where $T^A$ are the generators of the representation $R$. Due to the gauge invariance of the theory (\ref{Action}) the Yukawa couplings $\lambda_0$ should satisfy the equation

\begin{equation}\label{ConditionOnYukawa}
(T^A)_l{}^k \lambda_0^{ijl}  + (T^A)_l{}^j \lambda_0^{ilk}  + (T^A)_l{}^i \lambda_0^{ljk}  = 0.
\end{equation}

For investigating quantum corrections it is convenient to use the background field method \cite{DeWitt:1965jb,Abbott:1980hw,Abbott:1981ke}, which allows to construct the effective action invariant under the background gauge transformations. In the supersymmetric case \cite{Grisaru:1982zh,Gates:1983nr} the background-quantum splitting is non-linear and is performed by the substitution

\begin{equation}\label{BackgroundSubstitution}
e^{2V} \to e^{\bm{\Omega}^+} e^{2V} e^{\bm{\Omega}}
\end{equation}

\noindent
with background classical gauge superfield $\bm{V}$ defined as

\begin{equation}
e^{2\bm{V}}=e^{\bm{\Omega}^+}e^{\bm{\Omega}}.
\end{equation}

\noindent
After the replacement (\ref{BackgroundSubstitution}) the gauge superfield strength $W_a$ is split into the purely classical part and the part containing the background covariant derivatives acting on the superfield $V$,

\begin{equation}\label{NewW}
W_a \to \frac{1}{8} \bar{D}^2 (e^{-2\bm{V}} D_a e^{2\bm{V}}) + \frac{1}{8} e^{-\bm{\Omega}} \bm{\bar \nabla}^2 (e^{-2V} \bm{\nabla}_a e^{2V}) e^{\bm{\Omega}},
\end{equation}

\noindent
where

\begin{equation}
\bm{\nabla}_a = e^{-\bm{\Omega}^+} D_a e^{\bm{\Omega}^+}, \quad\bm{\bar \nabla}_{\dot{a}}=e^{\bm{\Omega}} \bar{D}_{\dot{a}} e^{-\bm{\Omega}}.
\end{equation}

\noindent
Due to the background gauge invariance it is possible to choose a gauge in which $\bm{\Omega}^+=\bm{\Omega}=\bm{V}$. Following Ref. \cite{Aleshin:2016yvj}, to regularize the theory, we add to the action a term $S_\Lambda$ containing higher covariant derivatives, such that

\begin{eqnarray}\label{Regularization}
&& S_{\mbox{\scriptsize reg}} \equiv S + S_{\Lambda} = \frac{1}{2e_0^2} \mbox{tr Re} \int d^4x\, d^2\theta\, e^{\bm{\Omega}} W^{a} e^{-\bm{\Omega}} R\Big( -\frac{\bar{\nabla}^2 \nabla^2}{16\Lambda^2} \Big)_{Adj} e^{\bm{\Omega}} W_{a} e^{-\bm{\Omega}}\nonumber\\
&& +\frac{1}{4}\int d^4x\, d^4\theta\, \phi^+ e^{\bm{\Omega}^+} e^{2V} F\Big(-\frac{\bar{\nabla}^2\nabla^2}{16\Lambda^2}\Big) e^{\bm{\Omega}}\phi + \left(\frac{1}{6}\int d^4x\,d^2\theta\, \lambda_0^{ijk}\phi_i\phi_j\phi_k + \mbox{c.c.}\right),\qquad
\end{eqnarray}

\noindent
where $W_a$ is given by Eq. (\ref{NewW}) and the subscript $Adj$ indicates that the covariant derivatives in the first term of Eq. (\ref{Regularization}) act on a superfield in the adjoint representation of the gauge group. The covariant derivatives entering Eq. (\ref{Regularization}) are defined as

\begin{equation}
\nabla_{a} = e^{-2V} \bm{\nabla}_{a} e^{2V}, \quad \bar{\nabla}_{\dot{a}} = \bm{\bar{\nabla}}_{\dot{a}}.
\end{equation}

\noindent
The parameter $\Lambda$ with the dimension of mass serves as an ultraviolet cut-off. The higher derivative regulator functions $R(y)$ and $F(y)$ should rapidly grow at infinity and satisfy the condition $R(0)=F(0)=1$.

The gauge-fixing term invariant under the background gauge transformations can be chosen in the form

\begin{equation}\label{Gauge-Fixing}
S_{\mbox{\scriptsize gf}} = -\frac{1}{16 e_0^2 \xi_0} \mbox{tr} \int d^4x\, d^4\theta\, \bm{\nabla}^2 V R\Big(-\frac{\bm{\bar{\nabla}}^2\bm{\nabla}^2}{16\Lambda^2}\Big)_{Adj} \bm{\bar{\nabla}}^2 V,
\end{equation}

\noindent
where $\xi_0$ is the bare gauge parameter. Although this term does not receive quantum corrections due to the Slavnov--Taylor identity \cite{Taylor:1971ff,Slavnov:1972fg}, the gauge parameter is renormalized even in the one-loop approximation \cite{Aleshin:2016yvj},

\begin{equation}
\frac{1}{\xi_0 e_0^2} = \frac{1}{\xi e^2} +\frac{C_2(1-\xi)}{12\pi^2\xi} \Big(\ln\frac{\Lambda}{\mu}+a_1\Big) + \mbox{higher orders},
\end{equation}

\noindent
where $e$ and $\xi$ are the renormalized gauge coupling constant and the renormalized gauge parameter, respectively. Here $\mu$ denotes the renormalization point and $a_1$ is a finite constant which depends on the subtraction scheme. In this paper for simplicity of calculations, we will use the Feynman gauge $\xi=1$, in which the one-loop correction vanishes, so that in the considered approximation it is possible to replace $e_0^2 \xi_0$ by $e^2$.

The gauge fixing procedure in the non-Abelian case also requires introducing the Faddeev--Popov and Nielsen--Kallosh ghosts. The Nielsen--Kallosh ghosts interact only with the background gauge superfield and contribute to the effective action only in the one-loop approximation. Because the diagrams considered in this paper do not contain ghost loops, we will not present explicit expressions for the ghost actions.

It is well known \cite{Faddeev:1980be} that the higher derivative terms do not remove one-loop divergences, so that it is necessary to insert into the generating functional the Pauli--Villars determinants \cite{Slavnov:1977zf}. To cancel the one-loop divergences (and subdivergences) coming from the loops of the quantum gauge superfield and of the ghosts, we introduce a set of three commuting chiral superfields in the adjoint representation. To cancel divergences coming from matter loops we introduce a set of massive anticommuting chiral superfields in the same representation of the gauge group as the matter superfields. However, diagrams considered in this paper do not include loops of the Pauli--Villars superfields, so that we will not discuss here the details, which can be found in Ref. \cite{Aleshin:2016yvj}. We only mention that the gauge fixed regularized theory is invariant under the BRST transformations \cite{Becchi:1974md,Tyutin:1975qk} which produce the Slavnov--Taylor identities at the quantum level.\footnote{The BRST transformations and the Slavnov--Taylor identities written in terms of superfields can be found in Ref. \cite{Ferrara:1975ye}.} Consequently, quantum corrections to the two-point Green function of the quantum gauge superfield $V$ are transversal,

\begin{equation}
\Gamma^{(2)}_{V}-S^{(2)}_{\mbox{\scriptsize gf}} = - \frac{1}{2e_0^2} \mbox{tr} \int\frac{d^4k}{(2\pi)^4} d^4\theta\, V(-k,\theta) \partial^2\Pi_{1/2} V(k,\theta)\, G_{V}(\alpha_0,\lambda_0,\Lambda/k).
\end{equation}

\noindent
(Note that in the tree approximation the function $G_V$ is equal to 1.) Due to the manifest background gauge invariance, the two-point Green function of the background superfield $\bm{V}$ is also transversal,

\begin{equation}
\Gamma^{(2)}_{\bm{V}} = -\frac{1}{8\pi} \mbox{tr} \int \frac{d^4p}{(2\pi)^4} d^4\theta\, \bm{V}(-p,\theta) \partial^2\Pi_{1/2} \bm{V}(p,\theta)\, d^{-1}(\alpha_0,\lambda_0, \Lambda/p),
\end{equation}

\noindent
where in the tree approximation $d^{-1} = \alpha_0^{-1}$ with $\alpha_0 = e_0^2/4\pi$. Also we will need the two-point Green function of the chiral matter superfields, for which the corresponding part of the effective action can be written as

\begin{equation}
\Gamma^{(2)}_{\phi} = \frac{1}{4}\int \frac{d^4q}{(2\pi)^4}d^4\theta\, \phi^{*i}(-q,\theta)\, \big(G_{\phi}\big)_i{}^j(\alpha_{0},\lambda_0,\Lambda/q)\, \phi_j(q,\theta).
\end{equation}

\noindent
The renormalized gauge coupling constant and the renormalization constants for the quantum gauge superfield and for the chiral matter superfields can be found from the requirement that the functions $d^{-1}$, $Z_V^2 G_V$,\footnote{In general, the quantum superfield $V$ is renormalized non-linearly \cite{Piguet:1981hh,Juer:1982fb,Juer:1982mp} (see also \cite{Piguet:1981mu,Piguet:1984mv}), but in this paper we need only the coefficient of the linear term, because in the considered approximation and for the considered terms non-linear renormalization is not essential.} and $(Z_\phi G_\phi)_i{}^j$ written in terms of the renormalized quantities should be finite in the limit $\Lambda\to \infty$. Note that due to the non-renormalization theorem for the superpotential \cite{Grisaru:1979wc}, the renormalization of the Yukawa couplings is related to the renormalization of the matter superfields,

\begin{equation}\label{YukawaRenorm}
\lambda^{ijk}=(\sqrt{Z_\phi})_l{}^{i}(\sqrt{Z_\phi})_m{}^j(\sqrt{Z_\phi})_n{}^k\lambda_0^{lmn}.
\end{equation}

RGFs defined in terms of the bare couplings can be obtained by differentiating the two-point Green functions with respect to $\ln\Lambda$ in the limit of the vanishing external momentum,

\begin{eqnarray}
&&\hspace*{-5mm} \frac{\beta(\alpha_0,\lambda_0)}{\alpha_0^2} = -\left.\frac{d\alpha_0^{-1}(\alpha,\lambda, \Lambda/\mu)}{d\ln\Lambda} \right|_{\alpha,\lambda=\mbox{\small const}} = \left.\frac{d}{d\ln\Lambda} (d^{-1}-\alpha_0^{-1}) \right|_{\alpha,\lambda=\mbox{\small const}, p\to 0};\label{BareBeta}\\
&&\hspace*{-5mm} \gamma_{V}(\alpha_0,\lambda_0) = -\left.\frac{d\ln Z_{V}(\alpha,\lambda,\Lambda/\mu)}{d\ln\Lambda} \right|_{\alpha,\lambda=\mbox{\small const}} = \frac{1}{2} \left.\frac{d \ln G_V(\alpha_0,\lambda_0,\Lambda/k)}{d\ln\Lambda} \right|_{\alpha,\lambda=\mbox{\small const}, k\to 0};\label{BareGammaV}\\
&&\hspace*{-5mm} (\gamma_\phi)_i{}^{j}(\alpha_0,\lambda_0) = -\left.\frac{d(\ln Z_\phi)_i{}^j(\alpha,\lambda,\Lambda/\mu)}{d\ln\Lambda} \right|_{\alpha,\lambda=\mbox{\small const}} = \left.\frac{d(\ln G_\phi)_i{}^{j}(\alpha_0,\lambda_0,\Lambda/q)}{d\ln\Lambda} \right|_{\alpha,\lambda=\mbox{\small const},q\to 0}\label{BareGammaPhi}.
\end{eqnarray}

\noindent
(The definition of the anomalous dimension $\gamma_c$ is not presented here, because in this paper we do not consider diagrams with the Faddeev--Popov ghost lines.) Note that  RGFs (\ref{BareBeta}) -- (\ref{BareGammaPhi}) depend on the regularization, but are scheme independent for a fixed regularization. In general, they differ from RGFs (standardly) defined in terms of the renormalized couplings,

\begin{eqnarray}\label{RenomalizedBeta}
&&\frac{\widetilde{\beta}(\alpha,\lambda)}{\alpha^2}=-\left.\frac{d\alpha^{-1}(\alpha_0,\lambda_0, \Lambda/\mu)}{d\ln\mu}\right|_{\alpha_0,\lambda_0=\mbox{\small const}};\\
\label{RenormalizedGammaV}
&&\widetilde{\gamma}_{V}(\alpha,\lambda)=\left.\frac{d\ln Z_{V}(\alpha,\lambda,\Lambda/\mu)}{d\ln\mu}\right|_{\alpha_0,\lambda_0=\mbox{\small const}};\\
\label{RenormalizedGammaPhi}
&&(\widetilde{\gamma}_\phi)_i{}^{j}(\alpha,\lambda)=\left.\frac{d(\ln Z_\phi)_i{}^j(\alpha,\lambda,\Lambda/\mu)}{d\ln\mu}\right|_{\alpha_0,\lambda_0=\mbox{\small const}},
\end{eqnarray}

\noindent
which are scheme and regularization dependent. Moreover, it is well known that RGFs (\ref{RenomalizedBeta}) -- (\ref{RenormalizedGammaPhi}) satisfy the NSVZ relation only in certain subtraction schemes.

\section{Dependence of the three-loop $\beta$-function on the Yukawa couplings}
\hspace*{\parindent}\label{SectionThreeLoopBeta}

Unlike Eq. (\ref{NSVZ}), the NSVZ relation in the form (\ref{RewrittenNSVZ}) (for RGFs defined in terms of the bare couplings) admits a simple graphical interpretation which is very similar to the one in the Abelian case.

Namely, let us consider a supergraph without external lines. By attaching to it two external lines of the background gauge superfield $\bm{V}$ we obtain superdiagrams that contribute to the $\beta$-function. By cutting its internal lines in all possible ways we obtain superdiagrams that contribute to the anomalous dimensions of the quantum superfields (i.e. of the quantum gauge superfield, of the Faddeev--Popov ghosts, and of the chiral matter superfields). The contribution to the beta-function and the contributions to the anomalous dimensions thus obtained are related to each other by the identity (\ref{RewrittenNSVZ}).

Note that the number of momentum integrations in the left hand side of Eq. (\ref{RewrittenNSVZ}) is equal to the number of loops $L$ in the original supergraph, while the number of momentum integrations in the right hand side is equal to $L-1$. Some explicit calculations made with the higher derivative regularization have demonstrated that the integrals in the left hand side are integrals of total derivatives \cite{Soloshenko:2003nc,Pimenov:2009hv} and even of double total derivatives \cite{Smilga:2004zr,Kazantsev:2014yna,Stepanyantz:2011bz,Shakhmanov:2017soc,Shakhmanov:2017wji}. Due to this feature it is possible to calculate one of the momentum integrals and obtain the same number of momentum integrations in both sides of Eq. (\ref{RewrittenNSVZ}) which is needed for the equality to take place.

\begin{figure}[h]
\begin{picture}(0,3)
\put(0.6,1.7){$(1)$}
\put(1.2,0.6){\includegraphics[scale=0.45]{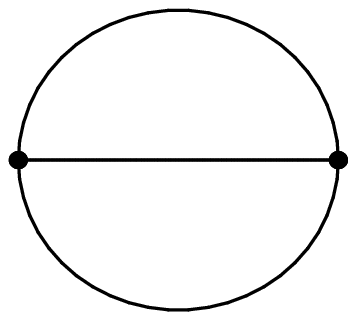}}
\put(3.6,1.7){$(2)$}
\put(4.2,0.6){\includegraphics[scale=0.45]{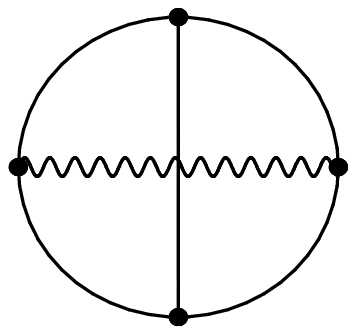}}
\put(6.6,1.7){$(3)$}
\put(7.2,0.6){\includegraphics[scale=0.45]{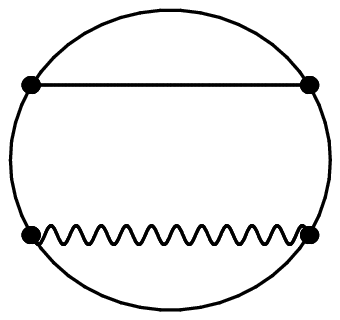}}
\put(9.6,1.7){$(4)$}
\put(10.2,0.1){\includegraphics[scale=0.45]{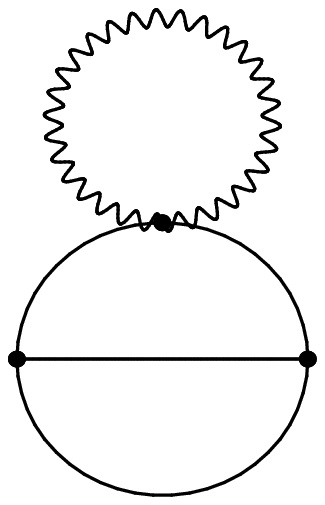}}
\put(12.6,1.7){$(5)$}
\put(13.2,0.58){\includegraphics[scale=0.45]{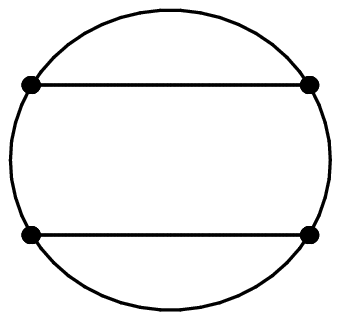}}
\end{picture}
\caption{Supergraphs generating the contributions to the $\beta$-function containing the Yukawa couplings in the three-loop approximation.}
\label{FigureBeta}
\end{figure}

In this section we calculate a part of the three-loop $\beta$-function containing the Yukawa couplings and compare it with the corresponding parts of the anomalous dimensions of the quantum superfields. The supergraphs generating the considered contributions are presented in Fig. \ref{FigureBeta}. The graphs (1) and (5) in Fig.~\ref{FigureBeta} have already been considered in Ref.~\cite{Shakhmanov:2017soc}, but, for the sake of completeness, we present the result for them below, together with the other contributions. Nevertheless, in this paper we will calculate the corresponding parts of RGFs for a more general form of the higher derivative regulator $F(y)$, than in Ref. \cite{Shakhmanov:2017soc}.

Below we present contributions corresponding to all graphs in Fig. \ref{FigureBeta} to the expression

\begin{equation}\label{WeCalculate}
\frac{d}{d\ln\Lambda}(d^{-1}-\alpha_0^{-1})\Bigr|_{\alpha,\lambda=\mbox{\small const};p\to 0}=\frac{\beta(\alpha_0,\lambda_0)}{\alpha_0^2}
\end{equation}

\noindent
in the form of the Euclidean momentum integrals,

\begin{eqnarray}
&&\hspace*{-9mm} \frac{d}{d\ln\Lambda}\mbox{graph}(1) = - \frac{2\pi}{r} C(R)_i{}^j \frac{d}{d\ln\Lambda} \int \frac{d^4k}{(2\pi)^4} \frac{d^4q}{(2\pi)^4} \lambda_0^{imn}\lambda^*_{0jmn}
\frac{\partial}{\partial q^\mu}\frac{\partial}{\partial q_{\mu}}\frac{1}{k^2F_k q^2 F_q (q+k)^2 F_{q+k}};\label{GraphOne}\nonumber\\
&&\\
&&\hspace*{-9mm}  \frac{d}{d\ln\Lambda}\mbox{graph}(2) = \frac{4\pi}{r} \frac{d}{d\ln\Lambda} \int\frac{d^4q}{(2\pi)^4}\frac{d^4l}{(2\pi)^4}\frac{d^4k}{(2\pi)^4}\, e_0^2 \Biggl[ \lambda^{*}_{0lkj} \lambda_{0}^{lki} C_{2} C(R)_{i}{}^{j} \frac{\partial}{\partial k^\mu}\left(\frac{\partial}{\partial k_\mu} -\frac{\partial}{\partial q_\mu}\right)\nonumber\\
&&\hspace*{-9mm}  -\Big( \lambda_{0jln}^{*} \lambda_{0}^{iln} \big(C(R)^2\big)_{i}{}^{j} - 2  \lambda^{*}_{0jln} \lambda_{0}^{imn} C(R)_{i}{}^{j} C(R)_{m}{}^{l} \Big) \frac{\partial}{\partial l^\mu}\left(\frac{\partial}{\partial l_\mu}+\frac{\partial}{\partial q_\mu}\right) +  \lambda_{0jln}^{*} \lambda_{0}^{iln} \big(C(R)^2\big)_{i}{}^{j} \nonumber\\
&&\hspace*{-9mm} \times \frac{\partial}{\partial q^\mu} \frac{\partial}{\partial q_\mu}\Biggr] \frac{N(q,k,l)}{k^2 R_{k} q^2 F_{q} (q+k)^2 F_{q+k} (q+k-l)^2 F_{q+k-l} (q-l)^2 F_{q-l} l^2 F_{l}};\label{GraphTwo}\\
&&\vphantom{1}\nonumber\\
&&\hspace*{-9mm} \frac{d}{d\ln\Lambda}\mbox{graph}(3) = -\frac{8\pi}{r} \frac{d}{d\ln\Lambda}\int\frac{d^4q}{(2\pi)^4}\frac{d^4l}{(2\pi)^4}\frac{d^4k}{(2\pi)^4}\, e_0^2\Biggl[ \lambda^{*}_{0lkj} \lambda_{0}^{lki} C_{2}C(R)_{i}{}^{j} \frac{\partial}{\partial k^\mu}\biggl(\frac{\partial}{\partial k_\mu}-\frac{\partial}{\partial q_\mu}\biggr)\nonumber\\
&&\hspace*{-9mm} + \lambda_{0jln}^{*} \lambda_{0}^{iln} \big(C(R)^2\big)_{i}{}^{j} \frac{\partial}{\partial q^\mu}\biggl(\frac{\partial}{\partial q_\mu}-\frac{\partial}{\partial l_\mu}\biggr) +  \lambda^{*}_{0jln} \lambda_{0}^{imn} C(R)_{i}{}^{j} C(R)_{m}{}^{l} \frac{\partial}{\partial l^\mu}\frac{\partial}{\partial l_\mu}\Biggr]\frac{L(q,q+k)}{k^2 R_k q^2 F_q^2 (q+l)^2}
\nonumber\\
&&\hspace*{-9mm} \times \frac{1}{F_{q+l} (q+k)^2 F_{q+k} l^2 F_l};\label{GraphThree}\\
&& \vphantom{1}\nonumber\\
&&\hspace*{-9mm} \frac{d}{d\ln\Lambda}\mbox{graph}(4) = \frac{8\pi}{r} \frac{d}{d\ln\Lambda} \int \frac{d^4q}{(2\pi)^4}\frac{d^4l}{(2\pi)^4} \frac{d^4k}{(2\pi)^4}\, e_0^2 \Biggl[ \lambda^{*}_{0lkj} \lambda_{0}^{lki} C_{2} C(R)_{i}{}^{j} \frac{\partial}{\partial k^\mu}\biggl(\frac{\partial}{\partial k_\mu}-\frac{\partial}{\partial q_\mu}\biggr)\nonumber\\
&&\hspace*{-9mm} + \lambda_{0jln}^{*} \lambda_{0}^{iln} \big(C(R)^2\big)_{i}{}^{j} \frac{\partial}{\partial q^\mu}\biggl(\frac{\partial}{\partial q_\mu}-\frac{\partial}{\partial l_\mu}\biggr) + \lambda^{*}_{0jln} \lambda_{0}^{imn} C(R)_{i}{}^{j} C(R)_{m}{}^{l} \frac{\partial}{\partial l^\mu}\frac{\partial}{\partial l_\mu}\Biggr]\frac{K(q,k)}{k^2 R_k q^2 F_q^2 l^2 F_l}\nonumber\\
&&\hspace*{-9mm} \times \frac{1}{(q+l)^2 F_{q+l}};\label{GraphFour}\\
&& \vphantom{1}\nonumber\\
&&\hspace*{-9mm} \frac{d}{d\ln\Lambda}\mbox{graph}(5) = \frac{4\pi}{r} C(R)_i{}^j \frac{d}{d\ln\Lambda} \int \frac{d^4k}{(2\pi)^4} \frac{d^4l}{(2\pi)^4} \frac{d^4q}{(2\pi)^4} \Biggl[\lambda_0^{iab} \lambda^*_{0kab} \lambda_0^{kcd} \lambda^*_{0jcd} \biggl(\frac{\partial}{\partial k^\mu} \frac{\partial}{\partial k_\mu} \nonumber\\
&&\hspace*{-9mm} -\frac{\partial}{\partial q^\mu} \frac{\partial}{\partial q_\mu}\biggr) + 2\lambda_0^{iab} \lambda_{0jac}^* \lambda_0^{cde} \lambda^*_{0bde} \frac{\partial}{\partial q^\mu}\frac{\partial}{\partial q_\mu}\Biggr] \frac{1}{k^2 F_k^2 q^2 F_q (q+k)^2 F_{q+k} l^2 F_l (l+k)^2 F_{l+k}}\label{GraphFive},
\end{eqnarray}

\noindent
where $R_k\equiv R(k^2/\Lambda^2)$ and $F_q\equiv F(q^2/\Lambda^2)$. Also Eqs. (\ref{GraphTwo}), (\ref{GraphThree}), and (\ref{GraphFour}) contain the functions $N(q,k,l)$, $L(q,q+k)$, and $K(q,k)$, which are defined by the equations

\begin{eqnarray}
&&\hspace*{-9mm} N(q,k,l) \equiv l^2 F_{q+k} F_{q+k-l} -(q-l)^2 \Big((q+k-l)^2-l^2\Big) F_{q+k} \frac{F_{q+k-l}-F_{q-l}}{(q+k-l)^2-(q-l)^2} \nonumber\\
&&\hspace*{-9mm} -q^2 \Big((q+k)^2 -l^2\Big) F_{q+k-l} \frac{F_{q+k}-F_{q}}{(q+k)^2-q^2} +q^2 (q-l)^2 \Big(l^2-(q+k)^2-(q+k-l)^2\Big) \nonumber\\
&&\hspace*{-9mm} \times \left(\frac{F_{q+k}-F_{q}}{(q+k)^2-q^2}\right) \left(\frac{F_{q+k-l}-F_{q-l}}{(q+k-l)^2-(q-l)^2}\right);\label{NFunction}\\
&&\hspace*{-9mm} L(q,p) \equiv F_q F_{p} + \frac{F_{p}-F_q}{p^2-q^2}\Big(F_q q^2 + F_p p^2 \Big) + 2 q^2 p^2 \left(\frac{F_p-F_q}{p^2-q^2}\right)^2;\label{LFunction}\\
&&\hspace*{-9mm} K(q,k) \equiv \frac{F_{q+k} - F_q - 2q^2 F_q'/\Lambda^2}{(q+k)^2 - q^2} + \frac{2q^2(F_{q+k}-F_q)}{\big((q+k)^2-q^2\big)^2}.\label{KFunction}
\end{eqnarray}

\noindent
(In our notation, the prime and the subscript $q$ denote the differentiation with respect to $q^2/\Lambda^2$.)

We see that all integrals (\ref{GraphOne}) -- (\ref{GraphFive}) are integrals of double total derivatives in the momentum space. This allows to calculate one of the loop integrals and to reduce thereby the number of the loop integrations by 1. Really, an integral of a total derivative can be transformed into a sum of surface integrals over an infinitely large sphere $S^3_\infty$ and over infinitely small spheres $S^3_{\varepsilon_i}$ surrounding singularities of the form $1/q_i^2$. Due to the higher derivative regularization the integral over $S^3_\infty$ vanishes, while integrals over $S^3_{\varepsilon_i}$ produce the result of the integration. For example, if $f(q^2)$ is a non-singular function which sufficiently rapidly decreases at infinity, then\footnote{Here we assume that the normal vector to the sphere $S^3_\varepsilon$ is inward-pointing.}

\begin{eqnarray}\label{IntegralOfTotalDerivative}
&& \int \frac{d^4q}{(2\pi)^4} \frac{\partial}{\partial q^\mu} \frac{\partial}{\partial q_\mu} \Big(\frac{f(q^2)}{q^2}\Big) = \frac{1}{16\pi^4} \int\limits_{S^3_\infty} dS_\mu\, \frac{\partial}{\partial q_\mu} \Big(\frac{f(q^2)}{q^2}\Big) + \frac{1}{16\pi^4} \int\limits_{S^3_\varepsilon} dS_\mu\, \frac{\partial}{\partial q_\mu} \Big(\frac{f(q^2)}{q^2}\Big)\qquad\nonumber\\
&& = \frac{1}{8\pi^4} \int\limits_{S^3_\varepsilon} dS\, \frac{f(q^2)}{q^3} = \frac{1}{4\pi^2} f(0).
\end{eqnarray}

\noindent
Calculating the integrals (\ref{GraphOne}) -- (\ref{GraphFive}) by this method, we obtain

\begin{eqnarray}\label{Graph1}
&&\hspace*{-6mm} \frac{d}{d\ln\Lambda}\mbox{graph}(1) = -\frac{1}{\pi r} C(R)_i{}^j \frac{d}{d\ln\Lambda} \int\frac{d^4k}{(2\pi)^4} \lambda_0^{imn} \lambda^*_{0jmn} \frac{1}{k^4 F_k^2};\\
&& \vphantom{1}\nonumber\\
\label{Graph2}
&&\hspace*{-6mm} \frac{d}{d\ln\Lambda}\mbox{graph}(2) = \frac{1}{\pi r}\frac{d}{d\ln\Lambda} \Bigg[ \int \frac{d^4q}{(2\pi)^4} \frac{d^4l}{(2\pi)^4}\, e_0^2\, \lambda^{*}_{0lkj} \lambda_{0}^{lki} C_{2} C(R)_{i}{}^{j} \frac{N(q,0,l)}{q^4 F_{q}^2 l^2 F_{l} (q-l)^4 F_{q-l}^2}\nonumber\\
&&\hspace*{-6mm} - \int\frac{d^4q}{(2\pi)^4}\frac{d^4k}{(2\pi)^4}\, e_0^2 \Big( \lambda_{0jln}^{*}\lambda_{0}^{iln} \big(C(R)^2\big)_{i}{}^{j} -2 \lambda^{*}_{0jln} \lambda_{0}^{imn} C(R)_{i}{}^{j} C(R)_{m}{}^{l}\Big) \frac{N(q,k,0)}{k^2 R_{k} q^4 F_{q}^2 (q+k)^4 F_{q+k}^2}\nonumber\\
&&\hspace*{-6mm} + 4 \int \frac{d^4k}{(2\pi)^4} \frac{d^4l}{(2\pi)^4}\, e_0^2\, \lambda_{0jln}^{*} \lambda_{0}^{iln} \big(C(R)^2\big)_{i}{}^{j} \frac{N(0,k,l)}{k^4 R_{k} F_{k} l^4 F_{l}^2 (k-l)^2 F_{k-l}}\Bigg];\\
&& \vphantom{1}\nonumber\\
\label{Graph3}
&&\hspace*{-6mm} \frac{d}{d\ln\Lambda}\mbox{graph}(3) = -\frac{2}{\pi r}\frac{d}{d\ln\Lambda} \Bigg[\int \frac{d^4q}{(2\pi)^4} \frac{d^4l}{(2\pi)^4}\,e_0^2\, \lambda^{*}_{0lkj} \lambda_{0}^{lki} C_{2} C(R)_{i}{}^{j} \frac{L(q,q)}{q^4 F_q^3 l^2 F_l (q+l)^2 F_{q+l}}\nonumber\\
&&\hspace*{-6mm} + \int \frac{d^4k}{(2\pi)^4} \frac{d^4l}{(2\pi)^4}\,e_0^2\, \lambda_{0jln}^{*} \lambda_{0}^{iln} \big(C(R)^2\big)_{i}{}^{j}  \left( \frac{L(0,k)}{k^4 R_k F_k l^4 F_l^2} +\frac{L(k,0)}{k^4 R_k F_k^2 l^2 F_l (l+k)^2 F_{l+k}}\right)\nonumber\\
&&\hspace*{-6mm} +2 \int\frac{d^4q}{(2\pi)^4}\frac{d^4k}{(2\pi)^4}\, e_0^2\, \lambda^{*}_{0jln} \lambda_{0}^{imn} C(R)_{i}{}^{j} C(R)_{m}{}^{l} \frac{L(q,q+k)}{k^2 R_k q^4 F_q^3 (q+k)^2 F_{q+k}}\Bigg];\\
&& \vphantom{1}\nonumber\\
\label{Graph4}
&&\hspace*{-6mm} \frac{d}{d\ln\Lambda}\mbox{graph}(4) = \frac{2}{\pi r} \frac{d}{d\ln\Lambda} \Bigg[\int\frac{d^4l}{(2\pi)^4}\frac{d^4q}{(2\pi)^4}\,e_0^2\, \lambda^{*}_{0lkj} \lambda_{0}^{lki} C_{2} C(R)_{i}{}^{j}\, \frac{K(q,0)}{q^2 F_q^2 l^2 F_l (q+l)^2 F_{q+l}}\nonumber\\
&&\hspace*{-6mm} + \int \frac{d^4k}{(2\pi)^4} \frac{d^4l}{(2\pi)^4}\,e_0^2\, \lambda_{0jln}^{*} \lambda_{0}^{iln} \big(C(R)^2\big)_{i}{}^{j}\, \frac{K(0,k)}{k^2R_k l^4 F_l^2}
+ \int\frac{d^4k}{(2\pi)^4} \frac{d^4q}{(2\pi)^4}\, e_0^2\, \lambda^{*}_{0jln} \lambda_{0}^{imn} C(R)_{i}{}^{j} C(R)_{m}{}^{l} \nonumber\\
&& \hspace*{-6mm} \times  \frac{2 K(q,k)}{k^2 R_k q^4 F_q^3}\Bigg];\\
&& \vphantom{1}\nonumber\\
\label{Graph5}
&&\hspace*{-6mm} \frac{d}{d\ln\Lambda} \mbox{graph}(5)=\frac{1}{\pi r} C(R)_i{}^j \frac{d}{d\ln\Lambda} \Bigg[ \lambda_0^{iab} \lambda^*_{0kab} \lambda_0^{kcd} \lambda_{0jcd}^* \int\frac{d^4k}{(2\pi)^4}\frac{d^4l}{(2\pi)^4}\frac{1}{k^4F_k^2l^4F_l^2}\nonumber\\
&&\hspace*{-6mm} +4 \lambda_0^{iab} \lambda_{0jac}^* \lambda_0^{cde} \lambda^*_{0bde} \int\frac{d^4k}{(2\pi)^4}\frac{d^4l}{(2\pi)^4}\frac{1}{k^4F_k^3l^2F_l(l+k)^2F_{l+k}}\Bigg].
\end{eqnarray}

These expressions contribute to the left-hand side of Eq. (\ref{RewrittenNSVZ}) written in terms of the bare couplings and should be compared with the corresponding contributions to the right hand side. To obtain them, we need to calculate the two-point Green functions for the chiral matter superfields and for the quantum gauge superfield $V$ in the two-loop approximation. This will be done in the next section.

\section{NSVZ relation for RGFs defined in terms of the bare couplings}
\hspace*{\parindent}\label{SectionBareNSVZ}

\begin{figure}[h]
\begin{picture}(0,12.3)
\put(0.4,10.6){\includegraphics[scale=0.4]{Yukawa2Loop}}
\put(2.25,11.05){\scalebox{1.5}{$\longrightarrow$}}
\put(3.5,11.7){\scriptsize{(1.1)}}\put(3.7,10.6){\includegraphics[scale=0.4]{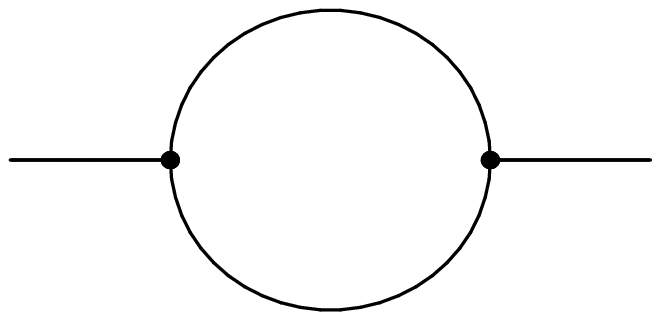}}

\put(0.4,8.6){\includegraphics[scale=0.4]{FirstYukawa3Loop}}
\put(2.25,9.08){\scalebox{1.5}{$\longrightarrow$}}
\put(3.5,9.7){\scriptsize{(2.1)}}\put(3.7,8.6){\includegraphics[scale=0.4]{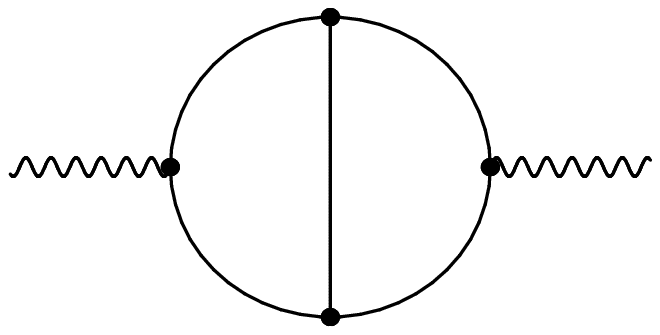}}
\put(6.6,9.7){\scriptsize{(2.2)}}\put(6.7,8.6){\includegraphics[scale=0.4]{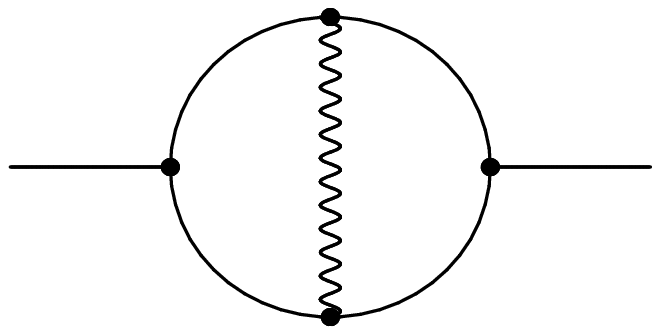}}
\put(9.6,9.7){\scriptsize{(2.3)}}\put(9.65,9.2){\includegraphics[scale=0.4]{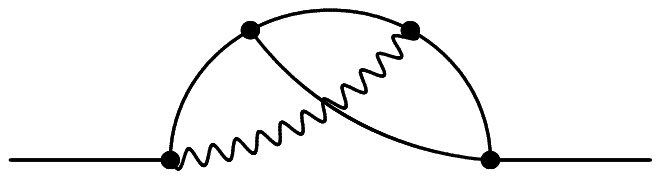}}

\put(0.4,5.6){\includegraphics[scale=0.4]{SecondYukawa3Loop}}
\put(2.25,6.05){\scalebox{1.5}{$\rightarrow$}}
\put(3.1,6.05){$\left\{\vphantom{\begin{array}{c}1\\0\\0\\0\\0\\0\\0\end{array}}\right.$}
\put(3.5,7.7){\scriptsize{(3.1)}}\put(3.7,6.63){\includegraphics[scale=0.4]{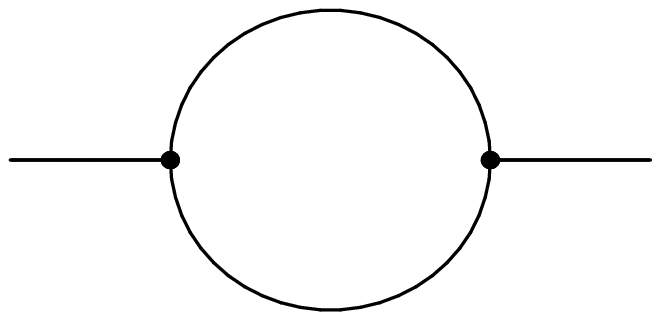}}
\put(6.6,7.7){\scriptsize{(3.2)}}\put(6.7,6.6){\includegraphics[scale=0.4]{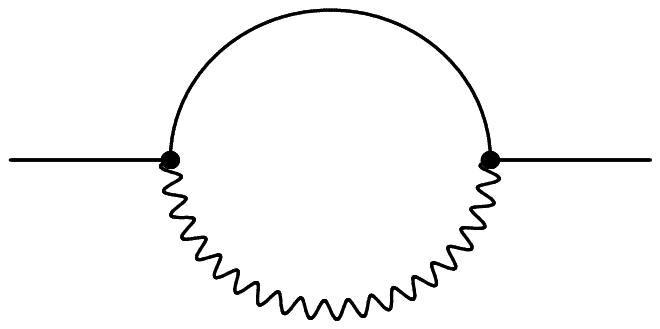}}
\put(9.6,7.7){\scriptsize{(3.3)}}\put(9.65,6.6){\includegraphics[scale=0.4]{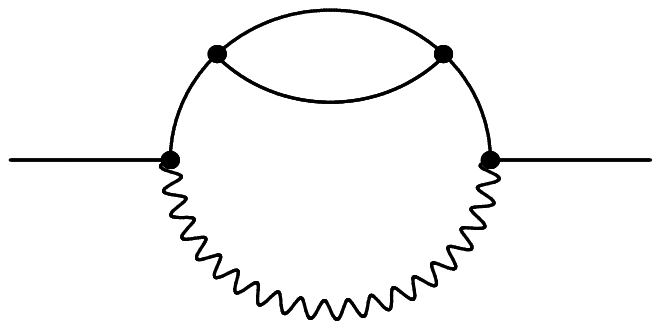}}
\put(3.5,5.7){\scriptsize{(3.4)}}\put(3.7,4.6){\includegraphics[scale=0.4]{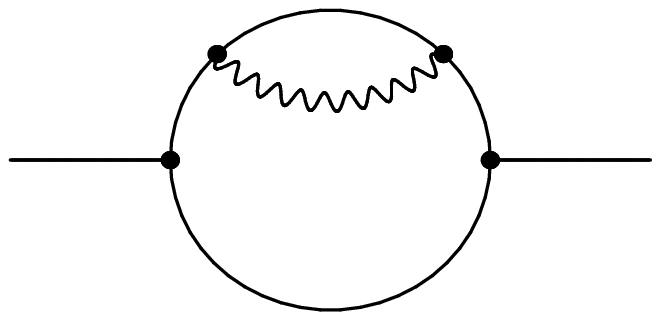}}
\put(6.6,5.7){\scriptsize{(3.5)}}\put(6.7,4.6){\includegraphics[scale=0.4]{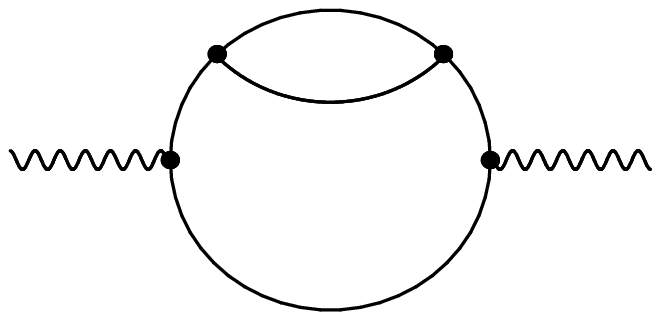}}

\put(0.4, 2.3){\includegraphics[scale=0.4]{ThirdYukawa3Loop}}
\put(2.25,2.95){\scalebox{1.5}{$\longrightarrow$}}
\put(3.5,3.6){\scriptsize{(4.1)}}\put(3.7,2.5){\includegraphics[scale=0.4]{ThirdYukawa3LoopGamma2}}
\put(6.6,3.6){\scriptsize{(4.2)}}\put(7.2,2.3){\includegraphics[scale=0.4]{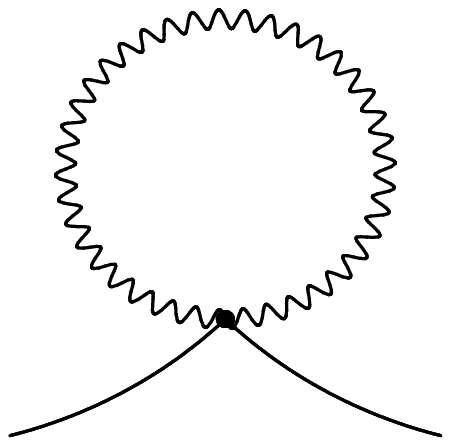}}
\put(9.6,3.6){\scriptsize{(4.3)}}\put(10.15,2.3){\includegraphics[scale=0.4]{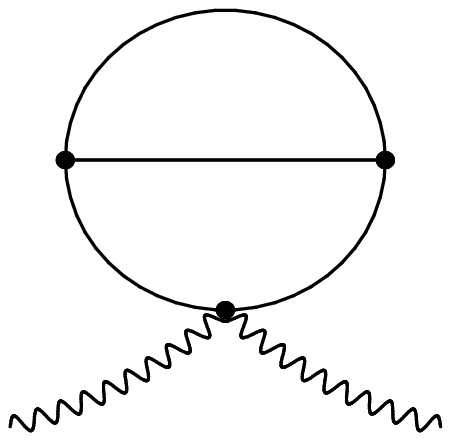}}
\put(12.7,3.6){\scriptsize{(4.4)}}\put(12.9,2.1){\includegraphics[scale=0.4]{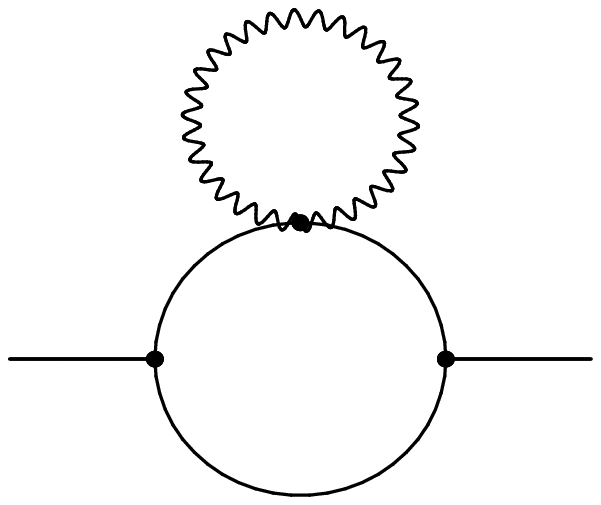}}

\put(0.4, 0.3){\includegraphics[scale=0.4]{FourthYukawa3Loop}}
\put(2.25,0.8){\scalebox{1.5}{$\longrightarrow$}}
\put(3.5,1.4){\scriptsize{(5.1)}}\put(3.7,0.35){\includegraphics[scale=0.4]{ThirdYukawa3LoopGamma2}}
\put(6.6,1.4){\scriptsize{(5.2)}}\put(6.8,0.35){\includegraphics[scale=0.4]{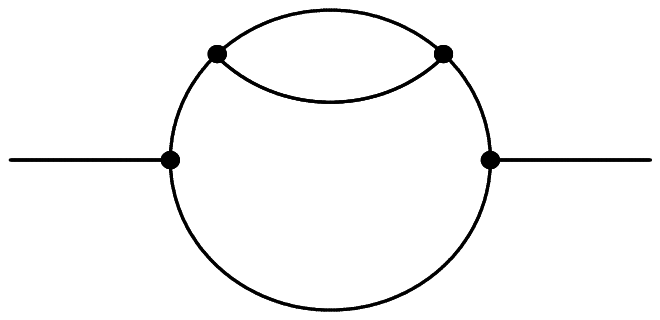}}

\end{picture}
\caption{Diagrams contributing to the two-point Green functions of the quantum gauge superfield $V$ and of the matter superfields, obtained from graphs in Fig.~\ref{FigureBeta} by cutting their internal lines.}
\label{FigureGamma}
\end{figure}

As we discussed above, to construct diagrams contributing to various anomalous dimensions which correspond to a certain graph in Fig.~\ref{FigureBeta}, it is necessary to cut internal lines in this graph in all possible ways and keep only 1PI superdiagrams. The result of this procedure is presented in Fig.~\ref{FigureGamma}. Note that some of the considered three-loop graphs also produce one-loop diagrams with two external matter lines. To understand, how the one-loop graphs contribute to the right-hand side of Eq. (\ref{RewrittenNSVZ}), we note that, according to Eq. (\ref{BareGammaPhi}), the anomalous dimension $(\gamma_\phi)_i{}^j$ is obtained by differentiating the function $(\ln G_\phi)_i{}^j$ with respect to $\ln\Lambda$. If we present the two-point Green function of the matter superfields in the form

\begin{equation}
(G_\phi)_i{}^j = \delta_i{}^j + (\Delta G_\phi)_i{}^j,
\end{equation}

\noindent
then

\begin{equation}\label{ExpansionG}
(\ln G_\phi)_i{}^j =\bigl(\ln(1+\Delta G_\phi)\bigr)_i{}^j =(\Delta G_\phi)_i{}^j - \frac{1}{2} (\Delta G_\phi)_i{}^k (\Delta G_\phi)_k{}^j +\dots
\end{equation}

\noindent
For the three-loop graphs (i.e. for the graphs (2) -- (5) in Fig. \ref{FigureBeta}) the first term corresponds to the two-loop diagrams, while the second term corresponds to products of the one-loop diagrams which are obtained by two cuts of internal lines. Really, the only possible way of cutting the graph (1) in Fig. 1 gives a single one-loop diagram, as is shown in Fig. \ref{FigureGamma}. It contributes to the first term in Eq. (\ref{ExpansionG}). However, the graphs (2) -- (5) can be cut in more than one way. A single cut of a matter line or of a gauge line gives a single two-loop diagram, while a double cut gives a pair of one-loop diagrams. The two-loop diagrams contribute to the first term in Eq. (\ref{ExpansionG}), while products of the pairs of one-loop diagrams contribute to the second term. Note that cutting the graph (5) twice, we obtain two copies of the same one-loop diagram. In this case its contribution to $(G_\phi)_i{}^j$, squared and multiplied by $1/2$, appears in the second term of Eq. (\ref{ExpansionG}). Also note that the labels of the diagrams presented in Fig. 2 are concordant to the labels of the graphs from which they have been obtained by cutting. Therefore, the same one-loop contribution

\begin{equation}
(\Delta G_\phi^{(1.1)})_i{}^j =(\Delta G_\phi^{(3.1)})_i{}^j = (\Delta G_\phi^{(4.1)})_i{}^j = (\Delta G_\phi^{(5.1)})_i{}^j
\end{equation}

\noindent
bears different labels for notational convenience.

The expressions obtained for all diagrams in Fig.~\ref{FigureGamma} are presented in Appendix A. Namely, in Appendix A we write their contributions to the functions $G_V(k)$ or $\big(G_\phi\big)_i{}^j(q)$ in the Euclidean space after the Wick rotation. These results should be compared with the expressions (\ref{Graph1}) -- (\ref{Graph5}), which encode contributions corresponding to the supergraphs presented in Fig. \ref{FigureBeta} to the function $\beta(\alpha_0,\lambda_0)/\alpha_0^2$,\footnote{Certainly, as we explained above, to obtain the contributions to the $\beta$-function, it is necessary to attach two external lines of the background gauge superfield $\bm{V}$ in all possible ways and sum expressions for all superdiagrams constructed in this way.}

\begin{equation}
\frac{d}{d\ln\Lambda}\mbox{graph}(A) \equiv \Delta_A \Big(\frac{\beta(\alpha_0,\lambda_0)}{\alpha_0^2}\Big),
\end{equation}

\noindent
where the subscript $A=1,\ldots, 5$ numerates the graphs in Fig. \ref{FigureBeta}. For each value of $A$ we have verified the equalities

\begin{equation}\label{RewrittenNSVZForGraphs}
\Delta_A \Big(\frac{\beta(\alpha_0,\lambda_0)}{\alpha_0^2}\Big) = \frac{1}{\pi} C_2 \Delta_A \gamma_V(\alpha_0,\lambda_0) - \frac{1}{2\pi r} C(R)_i{}^j (\Delta_A \gamma_\phi)_j{}^i(\alpha_0,\lambda_0),
\end{equation}

\noindent
where $\Delta_A \gamma_V(\alpha_0,\lambda_0)$ and $(\Delta_A \gamma_\phi)_j{}^i(\alpha_0,\lambda_0)$ are the contributions of the superdiagrams presented in Fig. \ref{FigureGamma} to the anomalous dimensions of the quantum gauge superfield and of the matter superfields, respectively,

\begin{eqnarray}
&& \Delta_1\gamma_V = 0;\qquad
\Delta_2\gamma_V = \frac{1}{2}\frac{d\Delta G_V^{(2.1)}}{d\ln\Lambda}\Big|_{k=0};\qquad
\Delta_3\gamma_V = \frac{1}{2}\frac{d\Delta G_V^{(3.5)}}{d\ln\Lambda}\Big|_{k=0};\qquad\nonumber\\
&& \Delta_4\gamma_V = \frac{1}{2}\frac{d\Delta G_V^{(4.3)}}{d\ln\Lambda}\Big|_{k=0};\qquad \Delta_5\gamma_V = 0;\\
&& \vphantom{1}\nonumber\\
\label{Delta1}
&& \big(\Delta_1 \gamma_\phi\big)_j{}^i =  \frac{d}{d\ln\Lambda}  \big(\Delta G_\phi^{(1.1)}\big)_j{}^i\Big|_{q=0}; \\
\label{Delta2}
&& \big(\Delta_2 \gamma_\phi\big)_j{}^i =  \frac{d}{d\ln\Lambda}\Big(\big(\Delta G_\phi^{(2.2)}\big)_j{}^i + \big(\Delta G_\phi^{(2.3)}\big)_j{}^i\Big)\Big|_{q=0};\\
\label{Delta3}
&& \big(\Delta_3 \gamma_\phi\big)_j{}^i =  \frac{d}{d\ln\Lambda} \Big(- \big(\Delta G_\phi^{(3.1)}\big)_j{}^k \big(\Delta G_\phi^{(3.2)}\big)_k{}^i + \big(\Delta G_\phi^{(3.3)}\big)_j{}^i + \big(\Delta G_\phi^{(3.4)}\big)_j{}^i\Big)\Big|_{q=0};\qquad\\
\label{Delta4}
&& \big(\Delta_4 \gamma_\phi\big)_j{}^i =  \frac{d}{d\ln\Lambda} \Big(- \big(\Delta G_\phi^{(4.1)}\big)_j{}^k \big(\Delta G_\phi^{(4.2)}\big)_k{}^i + \big(\Delta G_\phi^{(4.4)}\big)_j{}^i\Big)\Big|_{q=0};\\
\label{Delta5}
&& \big(\Delta_5 \gamma_\phi\big)_j{}^i =  \frac{d}{d\ln\Lambda}\Big(-\frac{1}{2} \big(\Delta G_\phi^{(5.1)}\big)_j{}^k \big(\Delta G_\phi^{(5.1)}\big)_k{}^i + \big(\Delta G_\phi^{(5.2)}\big)_j{}^i\Big)\Big|_{q=0}.
\end{eqnarray}

\noindent
To be more precise, in each of the five equalities (\ref{RewrittenNSVZForGraphs}) the integrals in the left hand side appear to be equal to the integrals in the right hand side. Using the one-loop expression for the $\beta$-function,

\begin{equation}
\frac{\beta(\alpha_0,\lambda_0)}{\alpha_0^2} = - \frac{1}{2\pi}\Big(3C_2-T(R)\Big) + O(\alpha_0,\lambda_0^2),
\end{equation}

\noindent
and taking into account that the considered groups of diagrams do not contribute to the anomalous dimension of the Faddeev--Popov ghosts $\gamma_c(\alpha_0,\lambda_0)$, we see that Eq. (\ref{RewrittenNSVZ}) is valid for the terms containing Yukawa couplings in the considered approximation. Moreover, due to Eq. (\ref{RewrittenNSVZForGraphs}) it is valid for each separate graph in Fig. \ref{FigureBeta}.

Note that due to the scheme-independence of RGFs defined in terms of the bare couplings, the NSVZ relation in the form (\ref{RewrittenNSVZ}) is valid in an arbitrary subtraction scheme in the case of using the higher covariant derivative regularization.

\section{RGFs defined in terms of the bare couplings}
\hspace*{\parindent}\label{SectionBareRGFs}

In the previous section we have verified that RGFs defined in terms of the bare couplings satisfy the NSVZ relation (\ref{RewrittenNSVZ}). Moreover, as we have demonstrated above, this equality takes place at the level of loop integrals. Therefore, for doing the verification described above, it is not necessary to calculate the integrals which give the anomalous dimensions of the quantum superfields. However, it would be desirable to obtain explicit expressions for these anomalous dimensions and for the $\beta$-function. This is done in this section for the higher derivative regulators

\begin{eqnarray}\label{RegulatorForm}
F(y)=1+y^n\qquad\mbox{and}\qquad R(y)=1+y^m.
\end{eqnarray}

\noindent
The detailed calculation of the anomalous dimensions is described in Appendix \ref{AppendixIntegrals}. Here we only present the main ideas and the results.

\subsection{Anomalous dimension of the quantum gauge superfield}
\hspace*{\parindent}

First, we find a part of the two-loop anomalous dimension of the quantum gauge superfield which contains the Yukawa couplings. It is given by the diagrams (2.1), (3.5), and (4.3),

\begin{equation}
\Delta\gamma_V = \sum\limits_{A=1}^5 \Delta_A\gamma_V = \frac{1}{2} \frac{d}{d\ln\Lambda}\Big(\Delta G_V^{(2.1)} + \Delta G_V^{(3.5)} + \Delta G_V^{(4.3)}\Big)\Big|_{k=0}.
\end{equation}

\noindent
This expression appears to be an integral of a double total derivative, which can be calculated by the help of Eq. (\ref{IntegralOfTotalDerivative}),

\begin{eqnarray}\label{DeltaGammaV}
&& \Delta\gamma_V(\alpha_0,\lambda_0) = - \frac{e_0^2}{4r} \lambda^{*}_{0jmn} \lambda_{0}^{imn} C(R)_{i}{}^{j} \frac{d}{d\ln\Lambda} \int\frac{d^4q}{(2\pi)^4}\frac{d^4l}{(2\pi)^4} \frac{\partial}{\partial q^\mu} \frac{\partial}{\partial q_\mu} \frac{1}{q^2 F_q (q+l)^2 F_{q+l} l^2 F_l}\qquad\nonumber\\
&& = -\frac{\alpha_0}{2\pi r} \lambda^{*}_{0jmn} \lambda_{0}^{imn} C(R)_{i}{}^{j} \frac{d}{d\ln\Lambda} \int\frac{d^4l}{(2\pi)^4} \frac{1}{l^4F_l^2} = -\frac{\alpha_0}{16\pi^3r} \lambda^{*}_{0jmn} \lambda_{0}^{imn} C(R)_{i}{}^{j}.
\end{eqnarray}

\noindent
Note that for terms of this structure we can ignore the dependence of the bare couplings on $\ln\Lambda$, because the one-loop renormalization of the gauge coupling constant does not include the Yukawa couplings,

\begin{equation}\label{AlphaRenorm}
\alpha^{-1} = \alpha_0^{-1} -\frac{1}{2\pi}\Big(3C_2 - T(R)\Big)\Big[\ln\frac{\Lambda}{\mu} + b_1\Big] + O(\alpha_0, \lambda_0^2).
\end{equation}

\noindent Taking into account the one-loop contribution found in Ref. \cite{Aleshin:2016yvj}, the considered part of the quantum gauge superfield anomalous dimension defined in terms of the bare couplings can be written as

\begin{equation}\label{GammaV}
\gamma_V(\alpha_0,\lambda_0) = -\frac{\alpha_0}{4\pi}\Big(3 C_2 - T(R)\Big) -\frac{\alpha_0}{16\pi^3r} \lambda^{*}_{0jmn} \lambda_{0}^{imn} C(R)_{i}{}^{j} + O(\alpha_0^2,\alpha_0\lambda_0^4).
\end{equation}

\noindent
This implies that in the considered approximation and for the terms of the considered structure

\begin{equation}
\gamma_V(\alpha_0,\lambda_0) = \frac{\beta(\alpha_0,\lambda_0)}{2\alpha_0} + O(\alpha_0^2,\alpha_0 \lambda_0^4).
\end{equation}

\noindent
Note that even in the one-loop approximation the calculation done in \cite{Aleshin:2016yvj} demonstrates that this equation is valid only in the Feynman gauge.

\subsection{Anomalous dimension of the chiral matter superfields}
\hspace*{\parindent}

The one- and two-loop contributions to the anomalous dimension of the chiral matter superfields of the considered structure can be written as

\begin{equation}
\big(\Delta\gamma_\phi\big)_i{}^j = \sum\limits_{A=1}^5 \big(\Delta_A \gamma_\phi\big)_i{}^j.
\end{equation}

\noindent
Adding the remaining part of the one-loop result (which does not contain the Yukawa couplings) we present the anomalous dimension in the form

\begin{equation}\label{GammaVOriginal}
\big(\gamma_\phi\big)_i{}^j = -\frac{\alpha_0}{\pi} C(R)_i{}^j + \sum\limits_{A=1}^5 \big(\Delta_A \gamma_\phi\big)_i{}^j + O(\alpha_0^2,\alpha_0\lambda_0^4,\lambda_0^6).
\end{equation}

\begin{figure}[h]
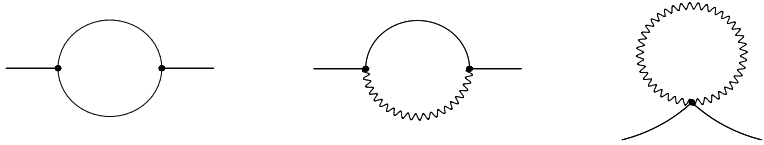

\begin{picture}(0,2.3)
\put(2.5,0.3){\includegraphics[scale=0.42]{SecondYukawa3LoopGamma1}}
\put(6.55,0.25){\includegraphics[scale=0.42]{SecondYukawa3LoopGamma2}}
\put(10.6,0.0){\includegraphics[scale=0.42]{ThirdYukawa3LoopGamma1}}
\end{picture}
\caption{One-loop graphs contributing to the renormalization of the chiral matter superfields.}
\label{OneLoopRenorm}
\end{figure}

\noindent
Calculating this expression it is necessary to take into account the one-loop renormalization of the Yukawa couplings. (Note that due to Eq. (\ref{AlphaRenorm}) renormalization of the gauge coupling constant is not essential in the considered approximation.) Due to Eq. (\ref{YukawaRenorm}) the renormalization of the Yukawa couplings is related to the renormalization of the chiral matter superfields. In the one-loop approximation it is defined by the diagrams presented in Fig. \ref{OneLoopRenorm}. With the considered regularization they have been calculated in \cite{Aleshin:2016yvj}. The result has the form

\begin{eqnarray}\label{GammaOneLoop}
&& (\gamma_{\phi})_i{}^j = \frac{d}{d\ln\Lambda} \int\frac{d^4k}{(2\pi)^4} \Big( - C(R)_i{}^j\frac{2 e_0^2}{k^4R_k}
+ \lambda^*_{0imn}\lambda_0^{jmn} \frac{2}{k^4 F_k^2}\Big) + O(\alpha_0^2,\alpha_0\lambda_0^2,\lambda_0^4)\qquad\nonumber\\
&& = -\frac{\alpha_0}{\pi}C(R)_i{}^j +\frac{1}{4\pi^2} \lambda^*_{0imn} \lambda_0^{jmn}  + O(\alpha_0^2,\alpha_0\lambda_0^2,\lambda_0^4).
\end{eqnarray}

\noindent
Integrating the renormalization group equation (\ref{BareGammaPhi}), we obtain the one-loop renormalization constant

\begin{equation}\label{OneLoopZed}
(Z_\phi)_i{}^j = \delta_i{}^j +\frac{\alpha}{\pi} \Big(\ln\frac{\Lambda}{\mu}+g_{11}\Big) C(R)_i{}^j
-\frac{1}{4\pi^2} \lambda^*_{imn} \lambda^{jmn} \Big(\ln\frac{\Lambda}{\mu}+g_{12}\Big) +O(\alpha^2,\alpha\lambda^2,\lambda^4),
\end{equation}

\noindent
where $g_{11}$ and $g_{12}$ are finite constants that depend on the choice of the subtraction scheme. Then, the one-loop renormalization of the Yukawa couplings can be found by the help of Eq.~(\ref{YukawaRenorm}),

\begin{eqnarray}\label{OneLoopLambda}
&&\hspace*{-8mm} \lambda_0^{ijk} = \lambda^{ijk} -\frac{\alpha}{2\pi} \left(C(R)_m{}^i \lambda^{mjk} +C(R)_m{}^j \lambda^{imk} +C(R)_m{}^k \lambda^{ijm}\right)
\Big(\ln\frac{\Lambda}{\mu} + g_{11}\Big) +\frac{1}{8\pi^2}\nonumber\\
&&\hspace*{-8mm} \times \left(\lambda^{ijm} \lambda^*_{mab} \lambda^{kab} +\lambda^{imk} \lambda^*_{mab} \lambda^{jab} +\lambda^{mjk} \lambda^*_{mab} \lambda^{iab}\right)
\Big(\ln\frac{\Lambda}{\mu} + g_{12}\Big) + O(\alpha^2\lambda,\alpha\lambda^3,\lambda^5).\qquad
\end{eqnarray}

\noindent
From this equation we obtain

\begin{eqnarray}\label{UsefulRelation}
&&\lambda^{*}_{0imn}\lambda_0^{jmn}=\lambda^*_{imn}\lambda^{jmn} -\frac{\alpha}{\pi} \Big(\ln\frac{\Lambda}{\mu} + g_{11}\Big) \Big( C(R)_i{}^l \lambda^*_{lmn} \lambda^{jmn}
+ 2 C(R)_m{}^l \lambda^*_{iln} \lambda^{jmn}\Big)\qquad\nonumber\\
&& +\frac{1}{4\pi^2}\Big(\ln\frac{\Lambda}{\mu} + g_{12}\Big) \Big(\lambda^*_{iab} \lambda^{kab} \lambda^*_{kcd} \lambda^{jcd} + 2\lambda^*_{iac} \lambda^{jab} \lambda^*_{bde} \lambda^{cde}\Big) + O(\alpha^2\lambda^2, \alpha\lambda^4, \lambda^6).
\end{eqnarray}

\noindent
The calculation of the anomalous dimension (\ref{GammaVOriginal}) is described in Appendix \ref{AppendixIntegrals}. The result is

\begin{eqnarray}\label{GammaBare}
&&(\Delta \gamma_\phi)_i{}^j(\alpha_0,\lambda_0) =  \frac{1}{4\pi^2}\lambda^*_{0imn}\lambda_0^{jmn} -\frac{\alpha_0}{8\pi^3}  \lambda^*_{0lmn} \lambda_0^{jmn} C(R)_i{}^l \Big(1-\frac{1}{n}\Big) +\frac{\alpha_0}{4\pi^3}  \lambda^*_{0imn} \lambda_0^{jml}\qquad\nonumber\\
&& \times C(R)_l{}^n \Big(1+\frac{1}{n}\Big) -\frac{1}{16\pi^4} \lambda^*_{0iac} \lambda_0^{jab} \lambda^*_{0bde} \lambda_0^{cde}.\qquad
\end{eqnarray}

\subsection{$\beta$-function}
\hspace*{\parindent}

To find the considered part of the $\beta$-function defined in terms of the bare couplings, we recall that it satisfies the NSVZ relation (\ref{RewrittenNSVZ}) and is, therefore, completely defined by the anomalous dimensions presented above,

\begin{equation}
\Delta\Big(\frac{\beta(\alpha_0,\lambda_0)}{\alpha_0^2}\Big) \equiv \sum\limits_{A=1}^5 \Delta_A \Big(\frac{\beta(\alpha_0,\lambda_0)}{\alpha_0^2}\Big) = \frac{1}{\pi} C_2 \Delta \gamma_V(\alpha_0,\lambda_0) - \frac{1}{2\pi r} C(R)_i{}^j \big(\Delta \gamma_\phi\big)_j{}^i(\alpha_0,\lambda_0),
\end{equation}

\noindent
where $\Delta\gamma_V$ and $(\Delta \gamma_\phi)_j{}^i$ are given by Eqs. (\ref{DeltaGammaV}) and (\ref{GammaBare}), respectively. After adding the one-loop contribution the expression for the $\beta$-function can be presented in the form

\begin{eqnarray}\label{BetaBare}
&&\frac{\beta(\alpha_0,\lambda_0)}{\alpha_0^2} = -\frac{1}{2\pi}\Big(3 C_2 - T(R)\Big) + \Delta\Big(\frac{\beta(\alpha_0,\lambda_0)}{\alpha_0^2}\Big)\nonumber\\
&&\qquad\qquad\qquad\quad\ \ + O(\alpha_0^2\lambda_0^2,\alpha_0\lambda_0^4,\lambda_0^6)+\mbox{terms without the Yukawa couplings}\qquad\vphantom{\frac{1}{2}}\nonumber\\
&& = -\frac{1}{2\pi}\Big(3 C_2 - T(R)\Big) -\frac{1}{2\pi r}C(R)_j{}^i\left(\frac{1}{4\pi^2}\lambda^*_{0imn}\lambda_0^{jmn}\right.
+ \frac{\alpha_0}{8\pi^3} \lambda^*_{0imn} \lambda_0^{jmn} C_2  \nonumber\\
&& -\frac{\alpha_0}{8\pi^3} \lambda^*_{0lmn}\lambda_0^{jmn} C(R)_i{}^l \Big(1-\frac{1}{n}\Big) +\frac{\alpha_0}{4\pi^3}  \lambda^*_{0imn} \lambda_0^{jml} C(R)_l{}^n \Big(1+\frac{1}{n}\Big) -\frac{1}{16\pi^4} \lambda^*_{0iac}\qquad\nonumber\\
&& \left. \times  \lambda_0^{jab} \lambda^*_{0bde} \lambda_0^{cde} \vphantom{\frac{1}{2}}\right)
+ O(\alpha_0^2\lambda_0^2,\alpha_0\lambda_0^4,\lambda_0^6)+\mbox{terms without the Yukawa couplings}.
\end{eqnarray}

\section{RGFs defined in terms of the renormalized couplings}
\hspace*{\parindent}\label{SectionRenormalizedRGFs}

In the previous section we have obtained expressions for RGFs (i.e. for the anomalous dimensions and for the $\beta$-function) defined in terms of the bare couplings. Because the regularization is fixed, they do not depend on the renormalization prescription. As we have seen earlier, for the considered higher derivative regularization they satisfy the relation (\ref{RewrittenNSVZ}) independently of the subtraction scheme. However, according to \cite{Kataev:2013eta,Kataev:2013csa}, it is necessary to distinguish between RGFs defined in terms of the bare couplings and the ones (standardly) defined in terms of the renormalized couplings. The latter RGFs are scheme dependent and satisfy the NSVZ relation only in special subtraction schemes. For the non-Abelian gauge theories regularized by higher covariant derivatives such a subtraction scheme can be fixed by the conditions \cite{Stepanyantz:2016gtk}

\begin{equation}\label{Prescription}
Z_\alpha(\alpha,\lambda, x_0)=1,\quad (Z_\phi)_i{}^j(\alpha,\lambda, x_0)=\delta_i{}^j, \quad Z_c(\alpha,\lambda,x_0)=1,\quad Z_V(\alpha,\lambda,x_0)=1,
\end{equation}

\noindent
where $x_0$ is a certain value of $\ln\Lambda/\mu$. For $x_0=0$ this implies that only powers of $\ln\Lambda/\mu$ are included into the renormalization constants, so that it is possible to call this scheme $\mbox{HD}+\mbox{MSL}$, i.e. the higher covariant derivative regularization supplemented by minimal subtractions of logarithms. In this section we calculate RGFs defined in terms of the renormalized couplings. The prescription (\ref{Prescription}) for constructing the NSVZ scheme is verified in Sect. \ref{SectionNSVZScheme}.

\subsection{Anomalous dimension of the quantum gauge superfield}
\hspace*{\parindent}

First, we calculate the anomalous dimension of the quantum gauge superfield $V$. For this purpose we rewrite the right hand side of Eq. (\ref{GammaV}) in terms of the renormalized couplings using Eqs. (\ref{AlphaRenorm}) and (\ref{OneLoopLambda}). Then, we integrate the renormalization group equation (\ref{BareGammaV}), which gives

\begin{eqnarray}
&&\hspace*{-7mm} \ln Z_V = \frac{\alpha}{4\pi} \Big(3C_2-T(R)\Big)\Big[\ln\frac{\Lambda}{\mu} + v_1\Big] +\frac{\alpha}{16\pi^3r} \lambda^{*}_{jmn} \lambda^{imn} C(R)_{i}{}^{j}\Big[\ln\frac{\Lambda}{\mu} + v_2\Big] + O(\alpha^2,\alpha\lambda^4)\nonumber\\
&&\hspace*{-7mm} = \frac{\alpha_0}{4\pi} \Big(3C_2-T(R)\Big)\Big[\ln\frac{\Lambda}{\mu} + v_1\Big] +\frac{\alpha_0}{16\pi^3r} \lambda^{*}_{0jmn} \lambda_0^{imn} C(R)_{i}{}^{j}\Big[\ln\frac{\Lambda}{\mu} + v_2\Big] + O(\alpha_0^2,\alpha_0\lambda_0^4),
\end{eqnarray}

\noindent
where $v_1$ and $v_2$ are finite constants which define the subtraction scheme in the considered approximation. Differentiating $\ln Z_V$ with respect to $\ln\mu$ we obtain

\begin{equation}
\widetilde{\gamma}_V(\alpha,\lambda) = \frac{d\ln Z_V}{d\ln\mu} = -\frac{\alpha_0}{4\pi} \Big(3C_2-T(R)\Big) -\frac{\alpha_0}{16\pi^3r} \lambda^{*}_{0jmn} \lambda_0^{imn} C(R)_{i}{}^{j} + O(\alpha_0^2,\alpha_0\lambda_0^4).
\end{equation}

\noindent
The required anomalous dimension is obtained after rewriting this expression in terms of the renormalized couplings by the help of Eqs. (\ref{AlphaRenorm}) and (\ref{OneLoopLambda}),

\begin{equation}\label{GammaVRenorm}
\widetilde{\gamma}_V(\alpha,\lambda)= -\frac{\alpha}{4\pi} \Big(3C_2-T(R)\Big) -\frac{\alpha}{16\pi^3r} \lambda^{*}_{jmn} \lambda^{imn} C(R)_{i}{}^{j} + O(\alpha^2,\alpha\lambda^4).
\end{equation}

\noindent
We see that no finite constants enter this equation, so that the considered part of the anomalous dimension is scheme independent.

\subsection{Anomalous dimension of the chiral matter superfields}
\hspace*{\parindent}

The anomalous dimension of the matter superfields defined in terms of the renormalized couplings is calculated similarly to the anomalous dimension of the quantum gauge superfield. Integrating the renormalization group equation (\ref{BareGammaPhi}), taking Eqs. (\ref{AlphaRenorm}) and (\ref{OneLoopLambda}) into account, we obtain the corresponding renormalization constant,

\begin{eqnarray}\label{LogZed}
&&(\ln Z_\phi)_i{}^j =\frac{\alpha}{\pi} C(R)_i{}^j \Big[ \ln\frac{\Lambda}{\mu} +g_{11} \Big] -\frac{1}{4\pi^2} \lambda^*_{imn} \lambda^{jmn} \Big[ \ln\frac{\Lambda}{\mu} +g_{12} \Big]\nonumber\\
&& +\frac{\alpha}{4\pi^3} \lambda^*_{lmn} \lambda^{jmn} C(R)_i{}^l \Big[ \frac{1}{2}\ln^2\frac{\Lambda}{\mu} +g_{11}\ln\frac{\Lambda}{\mu} +\frac{1}{2} \Big(1-\frac{1}{n}\Big) \ln\frac{\Lambda}{\mu} +g_{21}\Big]\nonumber\\
&& +\frac{\alpha}{2\pi^3} \lambda^*_{imn} \lambda^{jml} C(R)_l{}^n \Big[ \frac{1}{2} \ln^2\frac{\Lambda}{\mu} +g_{11} \ln\frac{\Lambda}{\mu} -\frac{1}{2} \Big(1+\frac{1}{n}\Big) \ln\frac{\Lambda}{\mu} +g_{22}\Big] \nonumber\\
&& -\frac{1}{8\pi^4} \lambda^*_{iac} \lambda^{jab} \lambda^*_{bde} \lambda^{cde} \Big[\frac{1}{2} \ln^2\frac{\Lambda}{\mu} +g_{12} \ln\frac{\Lambda}{\mu} -\frac{1}{2} \ln\frac{\Lambda}{\mu} +g_{23}\Big]\nonumber\\
&& -\frac{1}{16\pi^4} \lambda^*_{iab} \lambda^{kab} \lambda^*_{kcd} \lambda^{jcd} \Big[ \frac{1}{2} \ln^2\frac{\Lambda}{\mu} +g_{12} \ln\frac{\Lambda}{\mu} +g_{24}\Big] + O(\alpha^2,\alpha\lambda^4,\lambda^6),\qquad
\end{eqnarray}

\noindent
where $g_{11}$ and $g_{12}$ are the same constants as in Eq. (\ref{OneLoopLambda}). Also it is necessary to introduce new constants $g_{21}$, $g_{22}$, $g_{23}$, and $g_{24}$, which define a subtraction scheme in the two-loop approximation.

For obtaining the anomalous dimension defined in terms of the renormalized couplings, we differentiate this expression with respect to $\ln\mu$ at fixed values of the bare couplings. The result should be reexpressed in terms of the renormalized quantities by the help of Eqs. (\ref{AlphaRenorm}) and (\ref{OneLoopLambda}). This gives

\begin{eqnarray}\label{GammaRenorm}
&&\hspace*{-10mm} (\widetilde{\gamma}_\phi)_i{}^j(\alpha,\lambda) = - \frac{\alpha}{\pi} C(R)_i{}^j + \frac{1}{4\pi^2} \lambda^*_{imn} \lambda^{jmn}
+ \frac{\alpha}{4\pi^3} \lambda^*_{lmn} \lambda^{jmn} C(R)_i{}^l \Big[ g_{12}-g_{11}-\frac{1}{2}\Big(1-\frac{1}{n}\Big)\Big]\nonumber\\
&&\hspace*{-10mm} + \frac{\alpha}{2\pi^3} \lambda^*_{imn} \lambda^{jml} C(R)_l{}^n \Big[g_{12}-g_{11}+\frac{1}{2}\Big(1+\frac{1}{n}\Big)\Big]
- \frac{1}{16\pi^4} \lambda^*_{iac} \lambda^{jab} \lambda^*_{bde} \lambda^{cde} + O(\alpha^2, \alpha\lambda^4, \lambda^6).
\end{eqnarray}

\noindent
We see that the two-loop terms proportional to $\alpha\lambda^2$ are scheme dependent due to the presence of the finite constants $g_{11}$ and $g_{12}$, while the one-loop terms and the term proportional to $\lambda^4$ are scheme independent.

It is expedient to compare this expression with the one obtained in the $\overline{\mbox{DR}}$ scheme, which can be found in \cite{Jack:1996vg}. The notations of this paper and of Ref. \cite{Jack:1996vg} are related by the equations

\begin{equation}\label{Notations}
\alpha = \frac{g^2}{4\pi};\qquad \lambda^{ijk} = \frac{1}{2} Y^{ijk};\qquad (\widetilde{\gamma}_{\phi,\overline{\mbox{\tiny DR}}})_i{}^j(\alpha,\lambda) = 2\gamma^j{}_i(g, Y);\qquad
\widetilde\beta_{\overline{\mbox{\tiny DR}}}(\alpha,\lambda) = \frac{g}{2\pi} \beta(g,Y).
\end{equation}

\noindent
By the help of these equations we obtain that in our notation the anomalous dimension in the $\overline{\mbox{DR}}$ scheme has the form

\begin{eqnarray}\label{GammaRenormDR}
&& (\widetilde{\gamma}_{\phi,\overline{\mbox{\tiny DR}}})_i{}^j(\alpha,\lambda)
= - \frac{\alpha}{\pi} C(R)_i{}^j + \frac{1}{4\pi^2} \lambda^*_{imn} \lambda^{jmn}
- \frac{\alpha}{8\pi^3}C(R)_i{}^l \lambda^*_{lmn} \lambda^{mnj} \nonumber\\
&&\qquad\qquad\qquad\quad\ \, + \frac{\alpha}{4\pi^3} \lambda^*_{imn} \lambda^{jml} C(R)_l{}^n
- \frac{1}{16\pi^4} \lambda^{jab} \lambda^*_{iac} \lambda^{cde} \lambda^*_{bde} + O(\alpha^2, \alpha\lambda^4, \lambda^6).\qquad
\end{eqnarray}

\noindent
We see that Eq. (\ref{GammaRenorm}) coincides with this result for

\begin{equation}\label{GammaDR_FiniteConstants}
g_{12}-g_{11} + \frac{1}{2n} = 0.
\end{equation}

\noindent
This confirms correctness of Eq. (\ref{GammaRenorm}), if we take into account that a proper choice of a subtraction scheme can relate results obtained with different
regularizations.

\subsection{$\beta$-function}
\hspace*{\parindent}

To find the $\beta$-function defined in terms of the renormalized couplings, we start with integrating Eq. (\ref{BareBeta}), into which we substitute the expression (\ref{BetaBare}), with respect to $\ln\Lambda$. (Note that, for completeness, in subsequent expressions we also write the one-loop contribution, although it does not contain the Yukawa couplings.) The result can be presented in the form

\begin{eqnarray}\label{CouplingRenorm}
&& \frac{1}{\alpha}-\frac{1}{\alpha_0} = -\frac{1}{2\pi}\Big(3C_2 - T(R)\Big)\Big[\ln\frac{\Lambda}{\mu} + b_1 \Big]\nonumber\\
&& -\frac{1}{2\pi r} C(R)_j{}^i \left(\frac{1}{4\pi^2} \lambda^*_{imn} \lambda^{jmn} \Big[ \ln\frac{\Lambda}{\mu}+b_2 \Big]
+ \frac{\alpha}{8\pi^3} \lambda^*_{imn} \lambda^{jmn} C_2 \Big[ \ln\frac{\Lambda}{\mu}+b_{31}\Big]
\right.\nonumber\\
&& -\frac{\alpha}{4\pi^3} \lambda^*_{lmn} \lambda^{jmn} C(R)_i{}^l \Big[\frac{1}{2} \ln^2\frac{\Lambda}{\mu} +g_{11} \ln\frac{\Lambda}{\mu} +\frac{1}{2}\Big(1-\frac{1}{n}\Big) \ln\frac{\Lambda}{\mu}  +b_{32}\Big]
\nonumber\\
&& - \frac{\alpha}{2\pi^3} \lambda^*_{imn} \lambda^{jml} C(R)_l{}^n \Big[ \frac{1}{2} \ln^2\frac{\Lambda}{\mu} +g_{11} \ln\frac{\Lambda}{\mu} -\frac{1}{2}\Big(1+\frac{1}{n}\Big) \ln\frac{\Lambda}{\mu} +b_{33}\Big]
\nonumber\\
&& +\frac{1}{8\pi^4} \lambda^*_{iac} \lambda^{jab} \lambda^*_{bde} \lambda^{cde} \Big[ \frac{1}{2} \ln^2\frac{\Lambda}{\mu} +g_{12} \ln\frac{\Lambda}{\mu} -\frac{1}{2} \ln\frac{\Lambda}{\mu} +b_{34}\Big]
\nonumber\\
&&\left. + \frac{1}{16\pi^4} \lambda^*_{iab} \lambda^{kab} \lambda^*_{kcd} \lambda^{jcd} \Big[\frac{1}{2}
\ln^2\frac{\Lambda}{\mu} +g_{12} \ln\frac{\Lambda}{\mu} +b_{35}\Big] \right)\nonumber\\
&& + O(\alpha^2\lambda^2,\alpha\lambda^4,\lambda^6) + \mbox{terms without the Yukawa couplings},\vphantom{\frac{1}{2}}
\end{eqnarray}

\noindent
where the finite constants $g_{11}$ and $g_{12}$ come from the one-loop renormalization of the Yukawa couplings described by Eq. (\ref{OneLoopLambda}), and several new constants $b_1$, $b_2$ and $b_{3i}$, with $i=1,...,5$, define the renormalized gauge coupling constant in the one-, two-, and three-loop approximation, respectively. We rewrite the right hand side of this equation in terms of the bare couplings using Eqs. (\ref{AlphaRenorm}) and (\ref{UsefulRelation}). Differentiating the result with respect to $\ln\mu$ at fixed values of the bare couplings, we obtain the considered contribution to the $\beta$-function defined in terms of the renormalized couplings,

\begin{eqnarray}\label{BetaRenorm}
&&\hspace*{-7mm} \frac{\widetilde{\beta}(\alpha,\lambda)}{\alpha^2} = -\frac{1}{2\pi}\Big(3C_2 - T(R)\Big) -\frac{1}{2\pi r} C(R)_j{}^i \left(\frac{1}{4\pi^2} \lambda^*_{imn} \lambda^{jmn} + \frac{\alpha}{8\pi^3} \lambda^*_{imn} \lambda^{jmn} C_2  +\frac{\alpha}{4\pi^3} \right.\nonumber\\
&&\hspace*{-7mm} \times \lambda^*_{lmn} \lambda^{jmn} C(R)_i{}^l \Big[ b_2-g_{11}-\frac{1}{2}\Big(1-\frac{1}{n}\Big) \Big]
+\frac{\alpha}{2\pi^3} \lambda^*_{imn} \lambda^{jml} C(R)_l{}^n \Big[b_2 -g_{11} +\frac{1}{2} \Big(1+\frac{1}{n}\Big)\Big] \nonumber\\
&&\hspace*{-7mm} \left. -\frac{1}{8\pi^4} \lambda^*_{iac}  \lambda^{jab} \lambda^*_{bde} \lambda^{cde} \Big[ b_2 -g_{12} +\frac{1}{2}\, \Big]  +\frac{1}{16\pi^4} \lambda^*_{iab} \lambda^{kab} \lambda^*_{kcd} \lambda^{jcd} \Big[ g_{12}-b_2\Big]\right)  + O(\alpha^2\lambda^2,\alpha\lambda^4,\lambda^6)\nonumber\\
&&\hspace*{-7mm} + \mbox{terms without the Yukawa couplings}.\vphantom{\frac{1}{2}}
\end{eqnarray}

\noindent
The presence of the finite constants $g_{11}$, $g_{12}$, and $b_2$ indicates the scheme dependence of the result. However, the term containing the factor $\alpha\lambda^2 C_2$ does not contain any arbitrary finite constants and is, therefore, scheme independent.

Let us also present the corresponding expression obtained in the $\overline{\mbox{DR}}$ scheme in Ref. \cite{Jack:1996vg}. In our notation (see Eq. (\ref{Notations})) it can be written as

\begin{eqnarray}\label{BetaRenormDR}
&&\hspace*{-7mm} \frac{\widetilde{\beta}_{\overline{\mbox{\tiny DR}}}(\alpha,\lambda)}{\alpha^2} = -\frac{1}{2\pi}\Big(3C_2 - T(R)\Big)
-\frac{1}{2\pi r}C(R)_j{}^i\Big[\frac{1}{4\pi^2} \lambda^*_{imn} \lambda^{jmn} + \frac{\alpha}{8\pi^3}  \lambda^*_{imn} \lambda^{jmn} C_2
- \frac{\alpha}{16\pi^3} \nonumber\\
&&\hspace*{-7mm} \times \lambda^*_{lmn} \lambda^{jmn} C(R)_i{}^l
+\frac{3\alpha}{8\pi^3} \lambda^*_{imn} \lambda^{jml} C(R)_l{}^n - \frac{1}{64\pi^4} \lambda^*_{iab} \lambda^{kab} \lambda^*_{kcd} \lambda^{jcd}
-\frac{3}{32\pi^4} \lambda^*_{iac} \lambda^{jab} \lambda^*_{bde} \lambda^{cde} \Big]\nonumber\\
&&\hspace*{-7mm}  + O(\alpha^2\lambda^2,\alpha\lambda^4,\lambda^6) + \mbox{terms without the Yukawa couplings}.\vphantom{\frac{1}{2}}
\end{eqnarray}

\noindent
Comparing this expression with Eq. (\ref{BetaRenorm}), we see that the results for the considered part of the $\beta$-function coincide if the finite constants satisfy the
equations

\begin{equation}
g_{12} - b_2 = -\frac{1}{4};\qquad b_2 - g_{11} + \frac{1}{2n} = \frac{1}{4}.
\end{equation}

\noindent
Note that they do not contradict Eq. (\ref{GammaDR_FiniteConstants}). Therefore, our results agree with the ones of Ref. \cite{Jack:1996vg} up to the choice of
the subtraction scheme. This fact can be considered as a test of the calculation correctness. It is known \cite{Avdeev:1981ew,Jack:1996vg,Jack:1996cn,Jack:1998uj} that in the $\overline{\mbox{DR}}$ scheme RGFs do not satisfy the NSVZ relation. This can be easily verified comparing the expressions (\ref{GammaRenormDR}) and (\ref{BetaRenormDR}).

\section{NSVZ scheme and minimal subtractions of logarithms}
\hspace*{\parindent}\label{SectionNSVZScheme}

According to Ref. \cite{Stepanyantz:2016gtk}, in the non-Abelian case the NSVZ scheme is possibly obtained by imposing the boundary conditions (\ref{Prescription}) on the renormalization constants. In particular, the values of $g_{11}$, $g_{12}$, and $b_2$ can be fixed by the equations

\begin{equation}\label{Scheme}
(Z_\phi)_i{}^j(\alpha,\lambda,x_0)=\delta_i{}^j, \qquad Z_{\alpha}(\alpha,\lambda,x_0)=1.
\end{equation}

\noindent
The second equation can be equivalently rewritten as $\alpha(\alpha_0,\lambda_0,x_0) = \alpha_0$. Replacing $\ln\Lambda/\mu$ with $x_0$ in Eqs. (\ref{LogZed}) and (\ref{CouplingRenorm}), from Eq. (\ref{Scheme}) we obtain

\begin{equation}\label{TheGees}
g_{11} = -x_0;\qquad g_{12} = -x_0;\qquad b_2=-x_0.
\end{equation}

\noindent
(We do not present values of other finite constants, because they do not enter the expressions for RGFs in the considered approximation.) Note that for $x_0 = 0$ all finite constants vanish and, therefore, only powers of $\ln\Lambda/\mu$ are included into the renormalization constants. This corresponds to the $\mbox{HD}+\mbox{MSL}$ renormalization prescription. For $x_0\ne 0$ it is possible to absorb $x_0$ into the redefinition of the renormalization scale $\mu$, so that RGFs remain unchanged.

Substituting the finite constants (\ref{TheGees}) corresponding to the scheme (\ref{Prescription}) into Eqs. (\ref{BetaRenorm}), (\ref{GammaRenorm}), and (\ref{GammaVRenorm}) in the considered approximation we obtain RGFs in this scheme,

\begin{eqnarray}
&&\hspace*{-7mm} \frac{\widetilde{\beta}_{\mbox{\tiny HD}+\mbox{\tiny MSL}}(\alpha,\lambda)}{\alpha^2} = -\frac{1}{2\pi}\Big(3C_2 - T(R)\Big) - \frac{1}{2\pi r} C(R)_j{}^i \Big[
\frac{1}{4\pi^2} \lambda^*_{imn} \lambda^{jmn} + \frac{\alpha}{8\pi^3} \lambda^*_{imn} \lambda^{jmn} C_2\nonumber\\
&&\hspace*{-7mm}  -\frac{\alpha}{8\pi^3} \lambda^*_{lmn}\lambda^{jmn} C(R)_i{}^l \Big(1-\frac{1}{n}\Big) +\frac{\alpha}{4\pi^3} \lambda^*_{imn} \lambda^{jml} C(R)_l{}^n \Big(1+\frac{1}{n}\Big) - \frac{1}{16\pi^4} \lambda^*_{iac} \lambda^{jab} \lambda^*_{bde} \lambda^{cde} \Big]\nonumber\\
&&\hspace*{-7mm} + \mbox{terms without the Yukawa couplings} + O(\alpha^2\lambda^2,\alpha\lambda^4,\lambda^6);\vphantom{\frac{1}{2}}\\
&&\vphantom{1}\nonumber\\
&&\hspace*{-7mm} (\widetilde{\gamma}_{\phi,{\mbox{\tiny HD}+\mbox{\tiny MSL}}})_i{}^j(\alpha,\lambda) = - \frac{\alpha}{\pi} C(R)_i{}^j + \frac{1}{4\pi^2} \lambda^*_{imn} \lambda^{jmn}
- \frac{\alpha}{8\pi^3} \lambda^*_{lmn} \lambda^{jmn} C(R)_i{}^l \Big(1-\frac{1}{n}\Big)\nonumber\\
&&\hspace*{-7mm} + \frac{\alpha}{4\pi^3} \lambda^*_{imn} \lambda^{jml} C(R)_l{}^n \Big(1+\frac{1}{n}\Big)
- \frac{1}{16\pi^4} \lambda^*_{iac} \lambda^{jab} \lambda^*_{bde} \lambda^{cde} + O(\alpha^2, \alpha\lambda^4,\lambda^6);\\
&&\vphantom{1}\nonumber\\
&&\hspace*{-7mm} \widetilde{\gamma}_{V,{\mbox{\tiny HD}+\mbox{\tiny MSL}}}(\alpha,\lambda) = -\frac{\alpha}{4\pi}\Big(3C_2 - T(R)\Big) -\frac{\alpha}{16\pi^3r} \lambda^{*}_{jmn} \lambda^{imn} C(R)_{i}{}^{j} + O(\alpha^2,\alpha\lambda^4).
\end{eqnarray}

\noindent
Comparing these expressions we see that in the $\mbox{HD}+\mbox{MSL}$ scheme the NSVZ relation (\ref{RewrittenNSVZ}) is satisfied,

\begin{eqnarray}
&&\hspace*{-7mm} \frac{\widetilde{\beta}_{\mbox{\tiny HD}+\mbox{\tiny MSL}}(\alpha,\lambda)}{\alpha^2} = -\frac{1}{2\pi}\Big[3C_2 - T(R)
- 2 C_2 \widetilde{\gamma}_{c,{\mbox{\tiny HD}+\mbox{\tiny MSL}}}(\alpha,\lambda)
- 2 C_2 \widetilde{\gamma}_{V,{\mbox{\tiny HD}+\mbox{\tiny MSL}}}(\alpha,\lambda) + \frac{1}{r} C(R)_j{}^i \nonumber\\
&&\hspace*{-7mm} \times (\widetilde{\gamma}_{\phi,{\mbox{\tiny HD}+\mbox{\tiny MSL}}})_i{}^j(\alpha,\lambda)\Big] + O(\alpha^2\lambda^2,\alpha\lambda^4,\lambda^6) + \mbox{terms without the Yukawa couplings}.
\end{eqnarray}

\noindent
Note that the anomalous dimension of the Faddeev--Popov ghosts $\widetilde\gamma_c$ does not contribute to the terms of the considered structure, because in the two-loop approximation it does not depend on the Yukawa couplings.

Furthermore, we see that the NSVZ relation in the form (\ref{NSVZ}) is also valid for the considered terms,

\begin{eqnarray}
&&\ \frac{\widetilde{\beta}_{\mbox{\tiny HD}+\mbox{\tiny MSL}}(\alpha,\lambda)}{\alpha^2} = -\frac{3C_{2}-T(R)+C(R)_{j}{}^{i} (\widetilde{\gamma}_{\phi,{\mbox{\tiny HD}+\mbox{\tiny MSL}}})_i{}^j/r}{2\pi(1-C_{2}\alpha/(2\pi))}\nonumber\\
&&\qquad\qquad\qquad\qquad  + O(\alpha^2\lambda^2,\alpha\lambda^4,\lambda^6) + \mbox{terms without the Yukawa couplings}.\qquad
\end{eqnarray}

\noindent
Thus, our calculation supports the guess that the $\mbox{HD}+\mbox{MSL}$ prescription gives the NSVZ scheme in the non-Abelian case. This verification is highly non-trivial because in the considered approximation the scheme dependence is essential.

\section{Conclusion}
\hspace*{\parindent}

In this paper we verify the guess of Ref. \cite{Stepanyantz:2016gtk} that with the Slavnov higher derivative regularization the perturbative calculations in
${\cal N}=1$ supersymmetric non-Abelian gauge theories give the exact NSVZ $\beta$-function in the form (\ref{RewrittenNSVZ}). This new form relates the $\beta$-function to the anomalous dimensions of the quantum gauge superfield, of the Faddeev--Popov ghosts, and of the chiral matter superfields. It is therefore different from the original NSVZ relation and is obtained by the help of the non-renormalization theorem for the triple gauge-ghost vertices proved in \cite{Stepanyantz:2016gtk}. The arguments of Ref. \cite{Stepanyantz:2016gtk} suggest that the new NSVZ relation is valid in the subtraction scheme given by the $\mbox{HD}+\mbox{MSL}$ prescription, which means that the calculations are to be done with the higher covariant derivative regularization and only powers of $\ln\Lambda/\mu$ are to be included into the renormalization constants.

To check the new form of the NSVZ relation, we compare the part of the three-loop $\beta$-function containing the Yukawa couplings with the corresponding parts of the two-loop anomalous dimensions of the quantum gauge superfield and of the chiral matter superfields. This allows to verify not only the original form of the NSVZ relation (\ref{NSVZ}), but also the new one given by Eq.~(\ref{RewrittenNSVZ}). It is important that the considered terms are scheme dependent, so that the check of Eq.~(\ref{RewrittenNSVZ}) made in this paper is rather non-trivial.

The calculations are made using the BRST-invariant version of the higher covariant derivative regularization. First, we find the considered contribution to the $\beta$-function defined in terms of the bare couplings and demonstrate that it can be presented as a sum of integrals of double total derivatives. Such a structure allows to calculate one of the loop integrals and compare the result with the corresponding contributions to the anomalous dimensions of the quantum gauge superfield and of the matter superfields. This check demonstrates that the new form of the NSVZ relation (\ref{RewrittenNSVZ}) is really valid for the considered terms independently of a subtraction scheme if the RGFs are defined in terms of the bare couplings. Note that the equality takes place even if the considered contributions to RGFs are written in the form of loop integrals. These integrals have been calculated for the higher derivative regulators (\ref{RegulatorForm}). Using the result of this calculation we obtain (scheme-dependent) RGFs defined in terms of the renormalized couplings. They agree with the ones obtained in the $\overline{\mbox{DR}}$ scheme in Ref. \cite{Jack:1996vg} up to the choice of the subtraction scheme. This allows to verify the correctness of the calculations. However, unlike the case of using dimensional reduction, with the higher covariant derivative regularization the NSVZ scheme can be naturally constructed by the help of the $\mbox{HD}+\mbox{MSL}$ prescription. In this paper we have checked that this prescription really gives the NSVZ relation for the considered terms both in the form (\ref{NSVZ}) and in the form (\ref{RewrittenNSVZ}).

\section*{Acknowledgements}
\hspace*{\parindent}

The authors are very grateful to A.L.Kataev for valuable discussions.

The work of A.E.K. is supported by the Foundation for the advancement of theoretical physics ``BASIS'', grant No 17-11-120.

\appendix

\section{Contributions of various supergraphs to the two-point Green functions}
\hspace*{\parindent}\label{AppendixDiagrams}

In this section we list the results for the superdiagrams presented in Fig.~\ref{FigureGamma}. They contribute to the two-point Green functions either of the chiral matter superfields $\phi$ or of the quantum gauge superfield $V$. All expressions presented below are written in the Euclidean space after the Wick rotation and depend on the Euclidean momentum. Let us also recall that the functions $N$, $L$, and $K$, which enter the equations below, are given by Eqs. (\ref{NFunction}), (\ref{LFunction}), and (\ref{KFunction}), respectively, and the prime together with the subscript $q$ denotes the derivative with respect to $q^2/\Lambda^2$.

First, we write the expressions for diagrams contributing to the two-point Green function of the quantum gauge superfield $V$,

\begin{eqnarray}
&&\Delta G_V^{(2.1)}(k) = \frac{2}{r} e_{0}^2\,  \lambda^{*}_{0jmn} \lambda_{0}^{imn} C(R)_{i}{}^{j} \int\frac{d^4q}{(2\pi)^4} \frac{d^4l}{(2\pi)^4} \frac{N(q,k,l)}{q^2F_{q}(q+k)^2F_{q+k}(q+k-l)^2F_{q+k-l}}\qquad\nonumber\\
&&\hspace{2.5cm}\times\frac{1}{(q-l)^2F_{q-l}l^2F_{l}};\\
&&\Delta G_{V}^{(3.5)}(k) = -\frac{4}{r} e_0^2\, \lambda^{*}_{0jmn} \lambda_{0}^{imn} C(R)_{i}{}^{j} \int\frac{d^4q}{(2\pi)^4} \frac{d^4l}{(2\pi)^4} \frac{L(q+k,q)}{q^2 F_q^2 l^2 F_l (q+k)^2 F_{q+k} (q+l)^2 F_{q+l}};\nonumber\\
&&\vphantom{1}\\
&&\Delta G_V^{(4.3)}(k)=\frac{4}{r} e_0^2\, \lambda^{*}_{0jmn} \lambda_{0}^{imn} C(R)_{i}{}^{j} \int\frac{d^4q}{(2\pi)^4} \frac{d^4l}{(2\pi)^4} \frac{K(q,k)}{q^2 F_q^2 l^2 F_l (q+l)^2 F_{q+l}}.
\end{eqnarray}

Next, we present expressions for diagrams contributing to the two-point Green function of the matter superfields. Note that the one-loop diagrams (1.1), (3.1), (4.1), and (5.1) are the same. Constructing the function $(G_\phi)_i{}^j$ it is necessary to include in it only one contribution, say, $(\Delta G_\phi^{(1.1)})_i{}^j$. The diagrams (3.1), (4.1), and (5.1) appear in Fig. \ref{FigureGamma} because this contribution becomes essential in higher orders, when we calculate the function $(\ln G_\phi)_i{}^j$.

\begin{eqnarray}
&&(\Delta G_\phi^{(1.1)})_i{}^j(q) = 2 \lambda_{0imn}^* \lambda_0^{jmn} \int\frac{d^4l}{(2\pi)^4}\frac{1}{l^2 F_{l} (q+l)^2 F_{q+l}};\label{OneLoopFirst}\\
&&(\Delta G_\phi^{(2.2)})_i{}^j(q)=4e_{0}^2\,\lambda_{0imn}^{*} \lambda_{0}^{jlk}\, \big(T^{A}\big)_{l}{}^{m} \big(T^{A}\big)_{k}{}^{n} \int\frac{d^4k}{(2\pi)^4}\frac{d^4l}{(2\pi)^4}\frac{N(l,k,q)}{k^2 R_{k} l^2 F_{l} (l+k)^2 F_{l+k}}\nonumber\\
&&\hspace{3cm}\times\frac{1}{(l+k-q)^2F_{l+k-q}(q-l)^2F_{q-l}};\label{TwoLoopZero}\\
&&(\Delta G_\phi^{(2.3)})_i{}^j(q) = -4e_{0}^2\, \lambda_{0lmn}^{*}\lambda_{0}^{jmn} C(R)_i{}^l \int\frac{d^4k}{(2\pi)^4}\frac{d^4l}{(2\pi)^4}\frac{N(q,k,l)+N(-q-k,k,-l)}{k^2 R_{k} l^2 F_{l} (q+k)^2 F_{q+k}}\nonumber\\
&&\hspace{3cm}\times\frac{1}{(q+k-l)^2F_{q+k-l}(q-l)^2F_{q-l}};\label{TwoLoopOne}\\
&&(\Delta G_\phi^{(3.2)})_i{}^j(q) = -2e_0^2\, C(R)_i{}^j \int\frac{d^4k}{(2\pi)^4}\frac{L(q,q+k)}{k^2 R_k (q+k)^2 F_{q+k}};\label{OneLoopSecond}\\
&&(\Delta G_\phi^{(3.3)})_i{}^j(q) = 4 e_0^2\, \lambda^{*}_{0lmn}\lambda_0^{jmn} C(R)_i{}^l \int\frac{d^4k}{(2\pi)^4}\frac{d^4l}{(2\pi)^4}\frac{L(q,q+k)}{k^2 R_{k} l^2 F_l (q+k)^2 F_{q+k}^2}\nonumber\\
&&\hspace{3cm}\times\frac{1}{(q+k+l)^2F_{q+k+l}};\label{TwoLoopTwo}\\
&&(\Delta G_\phi^{(3.4)})_i{}^j(q) = 8 e_0^2\, \lambda_{0imn}^*\lambda_0^{jml} C(R)_l{}^n \int\frac{d^4k}{(2\pi)^4}\frac{d^4l}{(2\pi)^4}\frac{L(q+l+k,q+l)}{k^2 R_{k} l^2 F_{l} (q+l+k)^2 F_{q+l+k}}\nonumber\\
&&\hspace{3cm}\times\frac{1}{(q+l)^2 F_{q+l}^2};\label{TwoLoopThree}\\
&&(\Delta G_\phi^{(4.2)})_i{}^j(q) = 2e_0^2\, C(R)_i{}^j \int\frac{d^4k}{(2\pi)^4}\frac{K(q,k)}{k^2R_k};\label{OneLoopThird}\\
&&(\Delta G_\phi^{(4.4)})_i{}^j(q)=-8 e_0^2\, \lambda^*_{0imn} \lambda_0^{jml} C(R)_l{}^n \int\frac{d^4k}{(2\pi)^4}\frac{d^4l}{(2\pi)^4} \frac{K(l,k)}{l^2 F_l^2 k^2 R_k (q+l)^2 F_{q+l}}\label{TwoLoopFour};\\
&&(\Delta G_\phi^{(5.2)})_i{}^j(q) = - \lambda^*_{0iac} \lambda_0^{jab} \lambda_{0bde}^{*} \lambda_0^{cde} \int\frac{d^4k}{(2\pi)^4}\frac{d^4l}{(2\pi)^4}\frac{8}{k^2 F_k^2 l^2 F_l (k+q)^2 F_{k+q} (l+k)^2 F_{l+k}}.\qquad\nonumber\\
\label{TwoLoopFive}
\end{eqnarray}

\section{Calculation of the integrals}
\hspace*{\parindent}\label{AppendixIntegrals}

In this section we find explicit expressions for the considered contributions to the anomalous dimensions of the superfield $V$ and of the matter superfields $\phi_i$. They  come from the diagrams presented in Fig. \ref{FigureGamma}. The former anomalous dimension is calculated for arbitrary regulator functions $R(y)$ and $F(y)$, while the result for the latter one is valid only for $R(y)=1+y^m$ and $F(y)=1+y^n$.

\subsection{Anomalous dimension of the quantum gauge superfield}
\hspace*{\parindent}

In the one-loop approximation the anomalous dimension of the quantum gauge superfield (defined in terms of the bare couplings) was calculated in \cite{Aleshin:2016yvj}. In the Feynman gauge the result can be written as

\begin{equation}
\gamma_V(\alpha_0,\lambda_0) = -\frac{\alpha_0}{4\pi}\Big(3C_2-T(R)\Big) + O(\alpha_0^2,\alpha_0\lambda_0^2).
\end{equation}

\noindent
Taking into account that the one-loop renormalization of the gauge coupling constant does not involve the Yukawa couplings, we see that the terms in $\gamma_V$ proportional to $\alpha_0\lambda_0^2$ are given by the expression

\begin{equation}
\frac{1}{2}\, \lim\limits_{k\to 0}\, \frac{d}{d\ln\Lambda}\Big(\Delta G_V^{(2.1)}(k) + \Delta G_V^{(3.5)}(k) + \Delta G_V^{(4.3)}(k) \Big),
\end{equation}

\noindent
where various $\Delta G_V(k)$ can be found in Appendix \ref{AppendixDiagrams}. Strictly speaking, the derivative with respect to $\ln\Lambda$ in this equation should be calculated at fixed values of the renormalized couplings and the one-loop contribution should be also taken into account. However, it is easy to see that in the considered approximation (we calculate only two-loop terms containing Yukawa couplings) the dependence of the bare couplings on $\ln\Lambda$ is not essential. That is why the anomalous dimension of the quantum gauge superfield can be written in the form

\begin{equation}
\gamma_V(\alpha_0,\lambda_0) = -\frac{\alpha_0}{4\pi}\Big(3C_2-T(R)\Big) + \Delta\gamma_{V}(\alpha_0,\lambda_0) + O(\alpha_0^2, \alpha_0\lambda_0^4),
\end{equation}

\noindent
where

\begin{eqnarray}
&&\Delta\gamma_{V}(\alpha_0,\lambda_0) = \frac{e_{0}^2}{r}\, \lambda^{*}_{0jmn}\lambda_{0}^{imn} C(R)_{i}{}^{j} \frac{d}{d\ln\Lambda}  \int\frac{d^4q}{(2\pi)^4}\frac{d^4l}{(2\pi)^4}\Biggl[\frac{N(q,0,l)}{q^4 F_{q}^2 (q-l)^4 F_{q-l}^2l^2F_{l}} \qquad\nonumber\\
&& - \frac{2 L(q,q)}{q^4 F_q^3 l^2F_l(q+l)^2F_{q+l}} + \frac{2 K(q,0)}{q^2 F_q^2l^2F_l(q+l)^2F_{q+l}} \Biggr].
\end{eqnarray}

\noindent
Remarkably, after some transformations this integral can be presented as the integral of a double total derivative,

\begin{eqnarray}
-\frac{e_0^2}{4r}\, \lambda^{*}_{0jmn}\lambda_{0}^{imn} C(R)_{i}{}^{j} \frac{d}{d\ln\Lambda} \int\frac{d^4q}{(2\pi)^4}\frac{d^4l}{(2\pi)^4} \frac{\partial}{\partial q^\mu}\frac{\partial}{\partial q_\mu}\frac{1}{q^2F_q(q+l)^2F_{q+l}l^2F_l}.
\end{eqnarray}

\noindent
This allows to calculate it analytically for an arbitrary regulator function $F(y)$,

\begin{equation}\label{CalculationofGammaV}
\Delta\gamma_V(\alpha_0,\lambda_0) = -\frac{\alpha_0}{2\pi r} \lambda^{*}_{0jmn} \lambda_{0}^{imn} C(R)_{i}{}^{j} \frac{d}{d\ln\Lambda}\int\frac{d^4l}{(2\pi)^4}\frac{1}{l^4F_l^2} = -\frac{\alpha_0}{16\pi^3r} \lambda^{*}_{0jmn} \lambda_{0}^{imn} C(R)_{i}{}^{j}.
\end{equation}

\noindent
(Note that the factorization of the integrals giving $\gamma_V$ into integrals of double total derivatives in the Feynman gauge has been earlier found in the one-loop approximation in Ref. \cite{Aleshin:2016yvj}.)

Thus, the anomalous dimension of the quantum gauge superfield defined in terms of the bare couplings can finally be written as

\begin{equation}
\gamma_V(\alpha_0,\lambda_0) = -\frac{\alpha_0}{4\pi}\Big(3C_2-T(R)\Big) -\frac{\alpha_0}{16\pi^3r} \lambda^{*}_{0jmn} \lambda_{0}^{imn} C(R)_{i}{}^{j} + O(\alpha_0^2, \alpha_0 \lambda_0^4).
\end{equation}

\subsection{Anomalous dimension of the matter superfields}
\hspace*{\parindent}

For calculating the anomalous dimension of the chiral matter superfields (defined in terms of the bare couplings) according to Eq. (\ref{BareGammaPhi}) we differentiate $(\ln G_\phi)_i{}^j$ with respect to $\ln\Lambda$ and take the limit of the vanishing momentum. In this paper we are interested in the terms containing the Yukawa couplings in the two-loop approximation. It is easy to see that the sum of these terms and the one-loop result can be written in the form

\begin{eqnarray}\label{DetailedGammaPhi}
&& \big(\gamma_\phi\big)_i{}^j = \lim\limits_{q\to 0}\,\frac{d\big(\ln G_\phi\big)_i{}^j}{d\ln\Lambda} = \lim\limits_{q\to 0}\,\frac{d}{d\ln\Lambda}
\left(\big(\Delta G_\phi^{(1.1)}\big)_i{}^j + \big(\Delta G_\phi^{(3.2)}\big)_i{}^j + \big(\Delta G_\phi^{(4.2)}\big)_i{}^j\vphantom{\frac{1}{2}}\right.\nonumber\\
&& - \frac{1}{2} \big(\Delta G_\phi^{(1.1)}\big)_i{}^k \big(\Delta G_\phi^{(1.1)}\big)_k{}^j - \big(\Delta G_\phi^{(1.1)}\big)_i{}^k \big(\Delta G_\phi^{(3.2)}\big)_k{}^j - \big(\Delta G_\phi^{(1.1)}\big)_i{}^k \big(\Delta G_\phi^{(4.2)}\big)_k{}^j\nonumber\\
&&\left. + \big(\Delta G_\phi^{(2.2)}\big)_i{}^j + \big(\Delta G_\phi^{(2.3)}\big)_i{}^j + \big(\Delta G_\phi^{(3.3)}\big)_i{}^j + \big(\Delta G_\phi^{(3.4)}\big)_i{}^j + \big(\Delta G_\phi^{(4.4)}\big)_i{}^j + \big(\Delta G_\phi^{(5.2)}\big)_i{}^j\vphantom{\frac{1}{2}}\right)\qquad\nonumber\\
&& + \mbox{two-loop terms without Yukawa couplings} + \mbox{higher orders}.\vphantom{\frac{1}{2}}
\end{eqnarray}

\noindent
The first three terms in this expression correspond to the one-loop approximation,

\begin{eqnarray}
&& \lim\limits_{q\to 0}\,\frac{d}{d\ln\Lambda}
\left(\big(\Delta G_\phi^{(1.1)}\big)_i{}^j + \big(\Delta G_\phi^{(3.2)}\big)_i{}^j + \big(\Delta G_\phi^{(4.2)}\big)_i{}^j\vphantom{\frac{1}{2}}\right)\nonumber\\
&&\qquad\qquad\qquad\qquad = \frac{d}{d\ln\Lambda} \int \frac{d^4k}{(2\pi)^4}\Big(- C(R)_i{}^j \frac{2e_0^2}{k^4 R_k} + \lambda^*_{0imn}\lambda_0^{jmn} \frac{2}{k^4 F_k^2} \Big).\qquad
\end{eqnarray}

\noindent
It is important that the derivative with respect to $\ln\Lambda$ in this expression is calculated at fixed values of the renormalized coupling constant $\alpha=e^2/4\pi$ and the renormalized Yukawa couplings $\lambda^{ijk}$. This implies that we should also differentiate $\alpha_0$ and $\lambda_0^{ijk}$. However, in the one-loop approximation the renormalization of the gauge coupling constant is independent of the Yukawa couplings. Therefore, for analyzing terms containing the Yukawa couplings one should take into account only renormalization of $\lambda$ described by Eq. (\ref{OneLoopLambda}),

\begin{eqnarray}\label{OneLoopResult}
&&\hspace*{-6mm} \lim\limits_{q\to 0}\,\frac{d}{d\ln\Lambda}
\left(\big(\Delta G_\phi^{(1.1)}\big)_i{}^j + \big(\Delta G_\phi^{(3.2)}\big)_i{}^j + \big(\Delta G_\phi^{(4.2)}\big)_i{}^j\vphantom{\frac{1}{2}}\right) = - e^2 C(R)_i{}^j \frac{d}{d\ln\Lambda} \int \frac{d^4k}{(2\pi)^4} \frac{2}{k^4 R_k}\nonumber\\
&&\hspace*{-6mm} + \frac{d}{d\ln\Lambda} \int \frac{d^4k}{(2\pi)^4} \frac{2}{k^4 F_k^2} \left[\lambda^*_{imn}\lambda^{jmn}\vphantom{\frac{1}{2}}\right. - \frac{\alpha}{\pi}\Big(\ln\frac{\Lambda}{\mu} + g_{11}\Big)\Big(C(R)_i{}^l \lambda^*_{lmn} \lambda^{jmn} + 2 C(R)_m{}^l \lambda^*_{iln} \lambda^{jmn}\Big)\nonumber\\
&&\hspace*{-6mm} \left. + \frac{1}{4\pi^2}\Big(\ln\frac{\Lambda}{\mu} + g_{12}\Big)\Big(\lambda^*_{iab} \lambda^{kab} \lambda^*_{kcd} \lambda^{jcd} + 2 \lambda^*_{iac} \lambda^{jab} \lambda^*_{bde} \lambda^{cde}\Big)\right]+ O(\alpha^2,\alpha\lambda^4,\lambda^6),\qquad
\end{eqnarray}

\noindent
where $O(\alpha^2,\alpha\lambda^4,\lambda^6)$ encodes two-loop terms without the Yukawa couplings and terms corresponding to higher orders (which appear due to renormalization of couplings).

Next, we consider the terms proportional to $\alpha_0 \lambda_0^2$ in Eq. (\ref{DetailedGammaPhi}). (Note that in the considered approximation in these terms we can substitute $\alpha_0\lambda_0^2$ with $\alpha\lambda^2$, because the difference is essential only in higher orders.) As a starting point, we note that the contribution of the diagram (2.2) in Fig. \ref{FigureGamma} vanishes,

\begin{eqnarray}\label{Vanishing}
&& \lim\limits_{q\to 0}\frac{d}{d\ln\Lambda}(\Delta G_\phi^{(2.2)})_i{}^j = -8 e^2\, \lambda_{imn}^{*} \lambda^{jlk} \big(T^{A}\big)_{l}{}^{m} \big(T^{A}\big)_{k}{}^{n} \frac{d}{d\ln\Lambda} \int\frac{d^4l}{(2\pi)^4} \frac{d^4k}{(2\pi)^4}\qquad \nonumber\\
&& \times \left(F_{l+k} + l^2\frac{F_{l+k}-F_l}{(l+k)^2-l^2}\right) \left(\frac{F_{l+k}-F_l}{(l+k)^2-l^2}\right) \frac{1}{k^2R_kl^2F_l^2(l+k)^2F_{l+k}^2} = 0.\qquad
\end{eqnarray}

\noindent
The last equality follows from the fact that the integral is convergent in both the ultraviolet and infrared regions. This implies that it does not depend on $\Lambda$ and, consequently, its derivative with respect to $\ln\Lambda$ vanishes. The remaining terms in Eq. (\ref{DetailedGammaPhi}) proportional to $\alpha \lambda^2$ can be written as

\begin{eqnarray}\label{BareQuadratic}
&&\hspace*{-8mm} \lim\limits_{q\to 0}\,\frac{d}{d\ln\Lambda}
\left( - \big(\Delta G_\phi^{(1.1)}\big)_i{}^k \big(\Delta G_\phi^{(3.2)}\big)_k{}^j - \big(\Delta G_\phi^{(1.1)}\big)_i{}^k \big(\Delta G_\phi^{(4.2)}\big)_k{}^j
+ \big(\Delta G_\phi^{(2.3)}\big)_i{}^j + \big(\Delta G_\phi^{(3.3)}\big)_i{}^j \right.
\nonumber\\
&&\hspace*{-8mm} \left. + \big(\Delta G_\phi^{(3.4)}\big)_i{}^j + \big(\Delta G_\phi^{(4.4)}\big)_i{}^j \right) = 8e^2\, \lambda^*_{imn} \lambda^{jml} C(R)_l{}^n \frac{d}{d\ln\Lambda} \int\frac{d^4k}{(2\pi)^4} \frac{d^4l}{(2\pi)^4} \frac{1}{l^4F_l^2k^2R_k(l+k)^2} \nonumber\\
&&\hspace*{-8mm} + 4e^2\, \lambda^*_{lmn} \lambda^{jmn} C(R)_i{}^l \frac{d}{d\ln\Lambda} \int\frac{d^4k}{(2\pi)^4} \frac{d^4l}{(2\pi)^4} \Biggl[-\frac{1}{k^4R_kl^2F_l(l+k)^2F_{l+k}}+\frac{1}{k^4R_kl^4F_l^2}\Biggr]\nonumber\\
&&\hspace*{-8mm} + 8e^2\, \lambda^*_{imn}\lambda^{jml} C(R)_l{}^n \frac{d}{d\ln\Lambda} \int\frac{d^4k}{(2\pi)^4} \frac{d^4l}{(2\pi)^4} \Biggl[\frac{1}{l^2F_l^3k^2R_k(l+k)^2F_{l+k}}\frac{F_{l+k}-F_l}{(l+k)^2-l^2}\nonumber\\
&&\hspace*{-8mm} \times\biggl(F_l+2(l+k)^2\frac{F_{l+k}-F_l}{(l+k)^2-l^2}\biggr) -\frac{2}{l^2F_l^3k^2R_k \big((l+k)^2-l^2\big)} \biggl(\frac{F_{l+k}-F_l}{(l+k)^2-l^2} -\frac{F'_l}{\Lambda^2}\biggr)\Biggr].\qquad
\end{eqnarray}

\noindent
The derivative of the last integral with respect to $\ln\Lambda$ in this expression vanishes. Really, as earlier, the integral converges in both ultraviolet and infrared regions and, therefore, is a constant independent of $\Lambda$. Note that subdivergences are also absent.

The terms proportional to $\lambda^4$ have been considered in \cite{Shakhmanov:2017soc} for $F(y)=1+y$. For completeness, here we also present the corresponding integrals. However, in this paper they will be calculated for $F(y)=1+y^n$. Again, for these terms in the considered approximation we can ignore the difference between $\lambda_0$ and $\lambda$, so that they can be written as

\begin{eqnarray}\label{Quartic}
&& \lim\limits_{q\to 0}\,\frac{d}{d\ln\Lambda}
\left(- \frac{1}{2} \big(\Delta G_\phi^{(1.1)}\big)_i{}^k \big(\Delta G_\phi^{(1.1)}\big)_k{}^j + \big(\Delta G_\phi^{(5.2)}\big)_i{}^j\vphantom{\frac{1}{2}}\right)
= \frac{d}{d\ln\Lambda} \int\frac{d^4k}{(2\pi)^4}\frac{d^4l}{(2\pi)^4}\nonumber\\
&&\qquad\qquad\quad \times \Bigg[ - \lambda^*_{iab} \lambda^{kab} \lambda^*_{kcd} \lambda^{jcd} \frac{2}{k^4 F_k^2 l^4 F_l^2} - \lambda^*_{iac} \lambda^{jab} \lambda^*_{bde} \lambda^{cde} \frac{8}{k^4F_k^3l^2F_l(l+k)^2F_{l+k}}\Bigg].\qquad\quad
\end{eqnarray}

Thus, collecting the results (\ref{OneLoopResult}) -- (\ref{Quartic}) (and omitting the vanishing terms), we present the anomalous dimension in the form

\begin{eqnarray}
&& \big(\gamma_\phi\big)_i{}^j(\alpha_0,\lambda_0) = - 8\pi\alpha C(R)_i{}^j \frac{d}{d\ln\Lambda} \int \frac{d^4k}{(2\pi)^4} \frac{1}{k^4 R_k} + \lambda^*_{imn}\lambda^{jmn} \frac{d}{d\ln\Lambda} \int \frac{d^4k}{(2\pi)^4} \frac{2}{k^4 F_k^2}\nonumber\\
&& + 16\pi\alpha \lambda^*_{lmn} \lambda^{jmn} C(R)_i{}^l \frac{d}{d\ln\Lambda} \Biggl[\int\frac{d^4k}{(2\pi)^4} \frac{d^4l}{(2\pi)^4} \Big(\frac{1}{k^4R_kl^4F_l^2}-\frac{1}{k^4R_kl^2F_l(l+k)^2F_{l+k}}\Big)\nonumber\\
&& -\frac{1}{8\pi^2} \biggl(\ln\frac{\Lambda}{\mu}+g_{11}\biggr) \int\frac{d^4l}{(2\pi)^4}\frac{1}{l^4F_l^2}\Biggr] + 32\pi\alpha\lambda^*_{imn}\lambda^{jml}C(R)_l{}^n\frac{d}{d\ln\Lambda} \Biggl[ \int\frac{d^4l}{(2\pi)^4}\frac{d^4k}{(2\pi)^4}\nonumber\\
&& \times \frac{1}{l^4F_l^2k^2R_k(l+k)^2} - \frac{1}{8\pi^2}\biggl(\ln\frac{\Lambda}{\mu}+g_{11}\biggr)\int\frac{d^4l}{(2\pi)^4}\frac{1}{l^4F_l^2}\Biggr]
- 2 \lambda^*_{iab} \lambda^{kab} \lambda^*_{kcd} \lambda^{jcd} \frac{d}{d\ln\Lambda}\nonumber\\
&&\times \Biggl[\int\frac{d^4k}{(2\pi)^4}\frac{d^4l}{(2\pi)^4}\frac{1}{k^4F_k^2l^4F_l^2} - \frac{1}{4\pi^2} \biggl(\ln\frac{\Lambda}{\mu}+g_{12}\biggr) \int\frac{d^4k}{(2\pi)^4} \frac{1}{k^4F_k^2}\Biggr] - 8 \lambda^*_{iac} \lambda^{jab} \lambda^*_{bde} \lambda^{cde} \nonumber\\
&& \times \frac{d}{d\ln\Lambda} \Biggl[\int\frac{d^4k}{(2\pi)^4} \frac{d^4l}{(2\pi)^4} \frac{1}{k^4F_k^3l^2F_l(l+k)^2F_{l+k}} - \frac{1}{8\pi^2} \biggl(\ln\frac{\Lambda}{\mu}+g_{12}\biggr) \int\frac{d^4k}{(2\pi)^4}\frac{1}{k^4F_k^2}\Biggr]\qquad\nonumber\\
&& +\mbox{two-loop terms without Yukawa couplings} + \mbox{higher orders}.\vphantom{\frac{1}{2}}
\end{eqnarray}

\noindent
After some simple transformations it is possible to present this expression in the form

\begin{eqnarray}\label{Gamma_I}
&& \big(\gamma_\phi\big)_i{}^j(\alpha_0,\lambda_0) = - 8\pi\alpha C(R)_i{}^j I_1  + 2\lambda^*_{imn}\lambda^{jmn} I_2 + 16\pi\alpha \lambda^*_{lmn} \lambda^{jmn} C(R)_i{}^l \left(I_{7} - I_{4} \vphantom{\frac{1}{2}}\right.\nonumber\\
&&\left. - I_{8} + \frac{1}{8\pi^2} \Big(\ln\frac{\Lambda}{\mu} + g_{11}\Big) I_1\right) + 32\pi\alpha\lambda^*_{imn}\lambda^{jml}C(R)_l{}^n \left(I_{3} + I_{9} - \frac{1}{8\pi^2} \Big(\ln\frac{\Lambda}{\mu} + g_{11}\Big)\right. \nonumber\\
&& \left.\vphantom{\frac{1}{2}}\times I_2\right) - 2  \lambda^*_{iab} \lambda^{kab} \lambda^*_{kcd} \lambda^{jcd} I_{6}  - 8 \lambda^*_{iac} \lambda^{jab} \lambda^*_{bde} \lambda^{cde} \left(I_{5} + I_{9}- \frac{1}{8\pi^2} \Big(\ln\frac{\Lambda}{\mu} + g_{12}\Big) I_2 \right) \vphantom{\frac{1}{2}}\qquad\nonumber\\
&& +\mbox{two-loop terms without Yukawa couplings} + \mbox{higher orders},\vphantom{\frac{1}{2}}
\end{eqnarray}

\noindent
where the integrals $I_1$ -- $I_{9}$ are defined as follows:

\begin{eqnarray}
&& I_1 \equiv \frac{d}{d\ln\Lambda} \int \frac{d^4k}{(2\pi)^4} \frac{1}{k^4 R_k} = \frac{1}{8\pi^2};\\
&& I_2 \equiv \frac{d}{d\ln\Lambda} \int \frac{d^4k}{(2\pi)^4} \frac{1}{k^4 F_k^2} = \frac{1}{8\pi^2};\\
&& I_{3} \equiv \frac{d}{d\ln\Lambda} \int\frac{d^4l}{(2\pi)^4} \frac{1}{l^4F_l^2}\left(\int \frac{d^4k}{(2\pi)^4} \frac{1}{k^2R_k(l+k)^2} - \frac{1}{8\pi^2} \ln \frac{\Lambda}{l}\right);\\
&& I_{4}\equiv \frac{d}{d\ln\Lambda} \int\frac{d^4k}{(2\pi)^4} \frac{1}{k^4R_k}\left(\int \frac{d^4l}{(2\pi)^4} \frac{1}{l^2F_l(l+k)^2F_{l+k}} - \frac{1}{8\pi^2} \ln\frac{\Lambda}{k}\right);\\
&& I_{5} \equiv \frac{d}{d\ln\Lambda} \int\frac{d^4k}{(2\pi)^4} \frac{1}{k^4F_k^3}\left(\int \frac{d^4l}{(2\pi)^4} \frac{1}{l^2F_l(l+k)^2F_{l+k}} - \frac{1}{8\pi^2} F_k \ln \frac{\Lambda}{k} \right);\\
&& I_{6} \equiv \frac{d}{d\ln\Lambda} \Biggl[\int\frac{d^4k}{(2\pi)^4}\frac{d^4l}{(2\pi)^4}\frac{1}{k^4F_k^2l^4F_l^2} - \frac{1}{4\pi^2} \biggl(\ln\frac{\Lambda}{\mu}+g_{12}\biggr) \int\frac{d^4k}{(2\pi)^4} \frac{1}{k^4F_k^2}\Biggr];\qquad\\
&& I_{7} \equiv \frac{d}{d\ln\Lambda} \Biggl[\int\frac{d^4k}{(2\pi)^4} \frac{d^4l}{(2\pi)^4} \frac{1}{k^4R_kl^4F_l^2} -\frac{1}{8\pi^2} \Big(\ln\frac{\Lambda}{\mu}+g_{11}\Big) \int\frac{d^4k}{(2\pi)^4}\Big(\frac{1}{k^4F_k^2} +\frac{1}{k^4 R_k}\Big)\Biggr];\qquad\quad\\
&& I_{8} \equiv  \frac{1}{8\pi^2} \int\frac{d^4k}{(2\pi)^4} \frac{1}{k^4}\ln\frac{\Lambda}{k}\, \frac{d}{d\ln\Lambda} \frac{1}{R_k};\\
&& I_{9} \equiv  \frac{1}{8\pi^2} \int \frac{d^4k}{(2\pi)^4} \frac{1}{k^4} \ln \frac{\Lambda}{k}\, \frac{d}{d\ln\Lambda} \frac{1}{F_k^2}.
\end{eqnarray}

The first two equations follow from the identity

\begin{equation}\label{UsefulIdentity}
\frac{d}{d\ln\Lambda} \int\frac{d^4k}{(2\pi)^4} \frac{f(k/\Lambda)}{k^4} = \frac{1}{8\pi^2} f(0),
\end{equation}

\noindent
where $f(y)$ is a nonsingular function rapidly decreasing at infinity. Using this identity (and taking into account that $\lim\limits_{k\to0} k^{2n}\ln \Lambda/k = 0$) we also obtain

\begin{eqnarray}\label{I3}
&& I_3 = \lim\limits_{l\to 0}\, \frac{1}{8\pi^2} \left(\int \frac{d^4k}{(2\pi)^4} \frac{1}{k^2R_k(l+k)^2} - \frac{1}{8\pi^2} \ln\frac{\Lambda}{l}\right);\\
\label{I4}
&& I_4 = I_5 = \lim\limits_{k\to 0}\, \frac{1}{8\pi^2} \left(\int \frac{d^4l}{(2\pi)^4} \frac{1}{l^2F_l(l+k)^2F_{l+k}} - \frac{1}{8\pi^2} \ln\frac{\Lambda}{k}\right).\qquad
\end{eqnarray}

\noindent
The integral in Eq. (\ref{I3}) has been calculated in \cite{Soloshenko:2003nc,Soloshenko:2002np} for the regulator $R(y)=1+y^m$,

\begin{equation}
\int\frac{d^4k}{(2\pi)^4}\frac{1}{k^2(l+k)^2R_k}=\frac{1}{8\pi^2}\Big(\ln\frac{\Lambda}{l}+\frac{1}{2}+o(1)\Big),
\end{equation}

\noindent
where $o(1)$ denotes terms vanishing in the limit $l\to 0$. Therefore, for this regulator

\begin{equation}
I_3 = \frac{1}{128\pi^4}.
\end{equation}

To evaluate the integral (\ref{I4}), first, we note that

\begin{eqnarray}\label{Difference}
&&\hspace*{-5mm} \lim\limits_{k\to 0}\int\frac{d^4l}{(2\pi)^4}\left(\frac{1}{l^2F_l(l+k)^2F_{l+k}} - \frac{1}{l^2(l+k)^2F_l^2}\right) = \lim\limits_{k\to 0} \int\frac{d^4l}{(2\pi)^4} \frac{F_{l}-F_{l+k}}{l^2F_l^2 (l+k)^2F_{l+k}}\nonumber\\
&&\hspace*{-5mm} = \lim\limits_{k\to 0} \left( \int \frac{d^4l}{(2\pi)^4} \frac{F_{l}-F_{k+l}}{l^2-(k+l)^2} \frac{1}{F_l^2 F_{k+l} (k+l)^2} - \int \frac{d^4l}{(2\pi)^4} \frac{F_{l}-F_{k+l}}{l^2-(k+l)^2} \frac{1}{F_l^2 F_{k+l} l^2}\right) = 0,\qquad
\end{eqnarray}

\noindent
because the last two integrals in Eq. (\ref{Difference}) are convergent in the limit $k \to 0$. Therefore, the considered integral can be reduced to a simpler one,

\begin{equation}
I_4 = I_5 = \lim\limits_{k\to 0}\, \frac{1}{8\pi^2} \left(\int \frac{d^4l}{(2\pi)^4} \frac{1}{l^2F_l^2(l+k)^2} - \frac{1}{8\pi^2} \ln\frac{\Lambda}{k}\right),
\end{equation}

\noindent
which can be calculated using the technique of Refs. \cite{Soloshenko:2003nc,Soloshenko:2002np,Soloshenko:2003sx}. Namely, in the four-dimensional spherical coordinates after integrating over the angles by the help of the equation

\begin{equation}
\int\limits_{-1}^1 dx\, \frac{\sqrt{1-x^2}}{l^2 + 2kl x+ k^2} = \left\{\begin{array}{l}
\displaystyle{\frac{\pi}{2k^2}}\quad \mbox{for}\quad k\ge l;\\
\vphantom{1}\\
\displaystyle{\frac{\pi}{2l^2}}\quad \mbox{for}\quad l\ge k,
\end{array} \right.
\end{equation}

\noindent
we obtain

\begin{equation}
I_4 = I_5 = \frac{1}{64\pi^4}  \lim\limits_{k\to 0}\, \Biggl( \int\limits_0^k\frac{ldl}{k^2F_l^2}  + \int\limits_k^\infty \frac{dl}{l} \frac{1}{F_l^2} - \int\limits_k^\Lambda \frac{dl}{l}\Biggr).
\end{equation}

\noindent
After taking the limit the result can be presented in the form

\begin{equation}
I_4 = I_5 = \frac{1}{64\pi^4}\Biggl(\frac{1}{2}+\int\limits_0^
\Lambda dl\, \frac{1-F_l^2}{l F_l^2}+\int\limits_\Lambda^\infty\frac{dl}{lF_l^2}\Biggr).
\end{equation}

\noindent
Next, we make the substitution $x=l/\Lambda$ in the first integral and $x=\Lambda/l$ in the second one,

\begin{equation}
I_4 = I_5 = \frac{1}{64\pi^4}\Biggl(\frac{1}{2}-\int\limits_0^1
dx\, \frac{2x^{2n-1}}{(1+x^{2n})^2}\Biggr) = \frac{1}{128\pi^4}\Big(1-\frac{1}{n}\Big).
\end{equation}

The integral $I_6$ has been calculated for an arbitrary function $F(y)$ in \cite{Shakhmanov:2017soc},

\begin{equation}
I_6 = -\frac{1}{32\pi^4}\Big(\ln\frac{\Lambda}{\mu} + g_{12}\Big).
\end{equation}

\noindent
Here we will not present details of this calculation. Instead of this, we describe a similar calculation of the integral $I_7$. It can be rewritten as

\begin{eqnarray}
&& I_7 = \lim\limits_{p\to 0}\, \frac{d}{d\ln\Lambda}\Bigg[\left(\int \frac{d^4k}{(2\pi)^4} \frac{1}{k^2 (k+p)^2 R_k} -\frac{1}{8\pi^2}\Big(\ln\frac{\Lambda}{\mu} + g_{11}\Big) \right)\left(\int \frac{d^4l}{(2\pi)^4} \frac{1}{l^2 (l+p)^2 F_l^2}\right.\nonumber\\
&&\left. -\frac{1}{8\pi^2}\Big(\ln\frac{\Lambda}{\mu} + g_{11}\Big) \right) - \frac{1}{64\pi^4}\Big(\ln\frac{\Lambda}{\mu} + g_{11}\Big)^2 \Bigg] = -\frac{1}{32\pi^4} \Big(\ln\frac{\Lambda}{\mu} + g_{11}\Big),
\end{eqnarray}

\noindent
because

\begin{eqnarray}
&& \frac{d}{d\ln\Lambda} \left(\int \frac{d^4k}{(2\pi)^4} \frac{1}{k^2 (k+p)^2 R_k} -\frac{1}{8\pi^2}\Big(\ln\frac{\Lambda}{\mu} + g_{11}\Big) \right) = O\Big(\frac{p^2}{\Lambda^2}\Big);\qquad\\
&& \frac{d}{d\ln\Lambda} \left(\int \frac{d^4l}{(2\pi)^4} \frac{1}{l^2 (l+p)^2 F_l^2} -\frac{1}{8\pi^2}\Big(\ln\frac{\Lambda}{\mu} + g_{11}\Big) \right) = O\Big(\frac{p^2}{\Lambda^2}\Big).
\end{eqnarray}

The integrals $I_8$ and $I_9$ are calculated using the equation

\begin{equation}
\int \frac{d^4k}{(2\pi)^4} \frac{1}{k^4} \ln\frac{\Lambda}{k}\, \frac{df(k/\Lambda)}{d\ln\Lambda}  = - \int \frac{d^4k}{(2\pi)^4} \frac{1}{k^4} \ln\frac{\Lambda}{k}\, \frac{df(k/\Lambda)}{d\ln k}  = - \frac{1}{8\pi^2} \int\limits_0^\infty \frac{dk}{k}\, \ln\frac{\Lambda}{k}\, \frac{df(k/\Lambda)}{d\ln k}.
\end{equation}

\noindent
Then, making the substitution $x=\ln(k/\Lambda)$ we obtain

\begin{equation}
\int \frac{d^4k}{(2\pi)^4} \frac{1}{k^4} \ln\frac{\Lambda}{k}\, \frac{df(k/\Lambda)}{d\ln\Lambda} = \frac{1}{8\pi^2} \int\limits_{-\infty}^\infty dx\, x\, \frac{df(e^x)}{dx}.
\end{equation}

\noindent
Therefore, for $R(y)=1+y^m$, $F(y)=1+y^n$

\begin{eqnarray}
&& I_8 = \frac{1}{64\pi^4} \int\limits_{-\infty}^\infty dx\, x\, \frac{d}{dx}\left(\frac{1}{1+e^{2mx}}\right) = - \frac{1}{32\pi^4}\int\limits_{-\infty}^{\infty}dx \frac{m x}{(e^{mx}+e^{-mx})^2}=0;\\
&& I_9 = \frac{1}{64\pi^4} \int\limits_{-\infty}^\infty dx\,x\, \frac{d}{dx}\left(\smash{\frac{1}{\big(1+e^{2nx}\big)^2}}\vphantom{\frac{1}{2}}\right) = \frac{1}{128\pi^4 n}.
\end{eqnarray}

Substituting the values of all integrals into Eq. (\ref{Gamma_I}) we obtain the expression for the anomalous dimension

\begin{eqnarray}
&&\hspace*{-12mm} \big(\gamma_\phi\big)_i{}^j(\alpha_0,\lambda_0) = - \frac{\alpha}{\pi} C(R)_i{}^j + \frac{1}{4\pi^2} \lambda^*_{imn}\lambda^{jmn} - \frac{\alpha}{4\pi^3} \lambda^*_{lmn} \lambda^{jmn} C(R)_i{}^l \left( \Big(\ln\frac{\Lambda}{\mu} + g_{11}\Big)\right.\nonumber\\
&&\hspace*{-12mm}\left. + \frac{1}{2}\Big(1-\frac{1}{n}\Big) \right) - \frac{\alpha}{2\pi^3}\lambda^*_{imn}\lambda^{jml}C(R)_l{}^n \left(\Big(\ln\frac{\Lambda}{\mu} + g_{11}\Big) -\frac{1}{2}\Big(1+\frac{1}{n}\Big)\right)  + \frac{1}{16\pi^4} \lambda^*_{iab} \lambda^{kab} \nonumber\\
&&\hspace*{-12mm}\times \lambda^*_{kcd} \lambda^{jcd} \Big(\ln\frac{\Lambda}{\mu} + g_{12}\Big) + \frac{1}{8\pi^4} \lambda^*_{iac} \lambda^{jab} \lambda^*_{bde} \lambda^{cde} \left( \Big(\ln\frac{\Lambda}{\mu} + g_{12}\Big) -\frac{1}{2}\right) + O(\alpha^2, \alpha\lambda^4, \lambda^6).
\end{eqnarray}

\noindent
It is important that the result should be expressed in terms of the bare couplings $\alpha_0$ and $\lambda_0$ using Eq. (\ref{UsefulRelation}). (Let us recall that for the considered terms the renormalization of the gauge coupling constant is not essential.) Then the terms containing $\ln(\Lambda/\mu)$ disappear and the result takes the form

\begin{eqnarray}
&& \big(\gamma_\phi\big)_i{}^j(\alpha_0,\lambda_0) = - \frac{\alpha_0}{\pi} C(R)_i{}^j + \frac{1}{4\pi^2} \lambda^*_{0imn}\lambda_0^{jmn} - \frac{\alpha_0}{8\pi^3} \lambda^*_{0lmn} \lambda_0^{jmn} C(R)_i{}^l \Big(1-\frac{1}{n}\Big)\qquad\nonumber\\
&& + \frac{\alpha_0}{4\pi^3}\lambda^*_{0imn}\lambda_0^{jml}C(R)_l{}^n \Big(1+\frac{1}{n}\Big) - \frac{1}{16\pi^4} \lambda^*_{0iac} \lambda_0^{jab} \lambda^*_{0bde} \lambda_0^{cde} + O(\alpha_0^2, \alpha_0\lambda_0^4, \lambda_0^6).\vphantom{\frac{1}{2}}
\end{eqnarray}

\end{document}